\documentclass[letterpaper]{report}\usepackage[]{graphicx}\usepackage[]{color}
\makeatletter
\def\maxwidth{ %
  \ifdim\Gin@nat@width>\linewidth
    \linewidth
  \else
    \Gin@nat@width
  \fi
}
\makeatother

\definecolor{fgcolor}{rgb}{0.345, 0.345, 0.345}

\usepackage{framed}
\makeatletter
 {\par\unskip\endMakeFramed%
 \at@end@of@kframe}
\makeatother

\definecolor{shadecolor}{rgb}{.97, .97, .97}
\definecolor{messagecolor}{rgb}{0, 0, 0}
\definecolor{warningcolor}{rgb}{1, 0, 1}
\definecolor{errorcolor}{rgb}{1, 0, 0}

\usepackage{alltt}
\usepackage{graphicx}
\usepackage{paralist}
\usepackage{color}
\usepackage{amsmath}
\usepackage{amsfonts}
\usepackage{amssymb} 
\usepackage{adjustbox}
\usepackage{setspace}
\usepackage{caption} 
\usepackage{subcaption} 

\usepackage{natbib}
\setcitestyle{authoryear,open={(},close={)}} 
\usepackage{parskip}
\DeclareGraphicsExtensions{.pdf,.png,.jpg}

\def\var{\textup{Var}}

\IfFileExists{upquote.sty}{\usepackage{upquote}}{}
\begin{document}
\title{Anisotropic local constant smoothing for \\ change-point regression function estimation}
\author{John R.J. Thompson, W. John Braun}
\maketitle

\newpage
John R.J. Thompson  \emph{(Corresponding Author)}

Department of Mathematics

Wilfrid Laurier University

Waterloo, Ontario N2L 3C5

johnthompson@wlu.ca
\\[3\baselineskip]
W. John Braun

Department of Computer Science, Mathematics, Physics, and Statistics

The University of British Columbia, Okanagan

Kelowna, British Columbia V1V 1V7

Tel: (250) 807-8032

john.braun@ubc.ca

\doublespacing

\begin{abstract}
Understanding forest fire spread in any region of Canada is critical to promoting forest health, and protecting human life and infrastructure. Quantifying fire spread from noisy images, where regions of a fire are separated by change-point boundaries, is critical to faithfully estimating fire spread rates. In this research, we develop a statistically consistent smooth estimator that allows us to denoise fire spread imagery from micro-fire experiments. We develop an anisotropic smoothing method for change-point data that uses estimates of the underlying data generating process to inform smoothing. We show that the anisotropic local constant regression estimator is consistent with convergence rate $O\left(n^{-1/{(q+2)}}\right)$. We demonstrate its effectiveness on simulated one- and two-dimensional change-point data and fire spread imagery from micro-fire experiments.
  \\[3\baselineskip]
\end{abstract}

\newcounter{l1}

\section{Introduction}

\begin{figure}[h!]

{\centering \includegraphics[width=.8\textwidth]{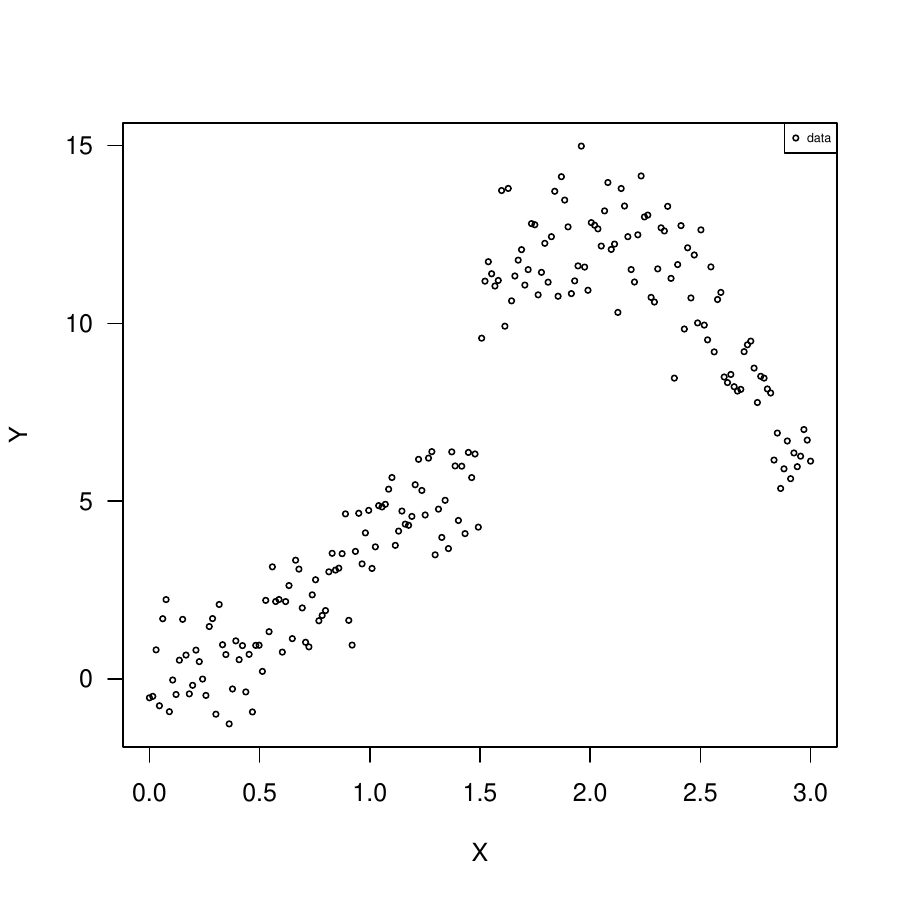} 

}

\caption[An example of one-dimensional change-point data, where there is an abrupt change in the regression function at $X=1.5.$]{An example of one-dimensional change-point data, where there is an abrupt change in the regression function at $X=1.5.$}\label{fig:exampleOfChangePoint}
\end{figure}

\begin{figure}[!htbp]
  \centering
  \includegraphics[width=0.7\textwidth]{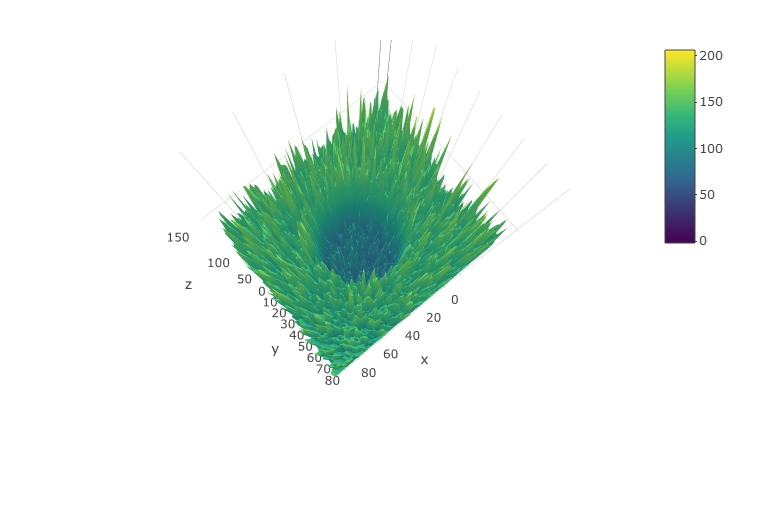}
\caption{An example of two-dimensional change-point data, where there is an abrupt change in the regression function at $X^2+Y^2=(20)^2$}
\label{fig:exampleOfChangePoint2D}
\end{figure}

Data that have an abrupt change in the regression function or its derivatives are classified as change-point data \citep{muller92}. An example of one-dimensional change-point data is shown in Figure \ref{fig:exampleOfChangePoint} where there is a clear abrupt jump in the regression function around $X=1.5$. Figure \ref{fig:exampleOfChangePoint2D} shows a two-dimensional change-``point"--a circular boundary--similar to the channels of imagery data.

Nonparametric kernel regression is a method of estimating a regression function without specifying the functional form of the relationship between regressors and outcome variables \citep{li07}. Kernel regression methods use local (about the regressor dimensions) information of the outcome to estimate the regression function and are known to oversmooth in regions of data where there is an abrupt change in the regression function when data, local in the $X$-direction, is not local in the $Y$-direction, or when $g'(x)/g(x)$ is very large \citep{fan96}. Popular kernel estimation methods are not well-equipped to differentiate between regions of the data that are separated by an abrupt change. For example, the Nadaraya-Watson kernel estimator inappropriately uses data from both sides of a change-point when estimating the regression function and oversmooths the regression function around change-points.

In image processing, an image filter is any technique that changes one or more characteristics of an image to create a new ``filtered" image. In this paper, we are interested in filters that remove undesirable noise in images. A popular method is Gaussian filtering, which is analogous to kernel smoothing. A Gaussian filter is in essence a local constant regression where the kernel is a mean-zero Gaussian distribution and the variance of the distribution acts as the smoothing parameter \citep{jain95}. The algorithm is a convolution between the Gaussian filter and the image \citep{nixon12} as
\begin{eqnarray*}\label{eqn:gaussianFilter}
 I(x,y,\sigma)=I(x,y,0)\circledast \frac{1}{2\pi\sigma^2}e^{-\left(\frac{x^2+y^2}{2\sigma^2}\right)},
\end{eqnarray*}
where $I(x,y,0)$ is red, green, blue (RGB) channel values at location (x,y) of the original image and $I(x,y,\sigma)$ is the convoluted image. The smoothing method of local constant kernel regression is equivalent to Gaussian filtering, where $h$ and $\sigma$ are analogous in controlling the smoothing. This method is also equivalent to the simplified heatflow diffusion \citep{nixon12} of
\begin{eqnarray} \label{eqn:diffusion}
  \frac{\partial I(x,y,t)}{\partial t}=\nabla^2 I(x,y,t),
\end{eqnarray}
where $I(x,y,t)$ is the RGB value of a pixel at coordinates $(x,y)$ and time $t$, $\nabla$ is the spatial gradient operator, and the initial condition $I(x,y,0)$ .

The specific area of image processing we are interested in is concerned with distinguishing boundaries or ``edges" in an image, and smoothing between the boundaries. Anisotropic diffusion filtering, also called Perona-Malik diffusion \citep{perona90,black98} or anisotropic diffusion, is a partial differential equation (PDE) method for edge-preserve smoothing between boundaries. Anisotropic diffusion is an edge-preserving extension to (isotropic) Gaussian filtering. A physical interpretation of anisotropic diffusion is that it takes a convolution of the image and an isotropic kernel function, using an edge-stopping function to preserve boundaries between regions without supervision. Essentially, they are taking locally weighted averages around each pixel. This is modeled analogously by a more general heatflow diffusion PDE given as
\begin{eqnarray}\label{eq:fullpde}
  \frac{\partial I(x,y,t)}{\partial t}=\nabla\cdot\left[f(\|\nabla I\|)\nabla I\right],
\end{eqnarray}
where $\nabla\cdot$ is the divergence operator, the edge-stopping function is $f(\|\nabla I\|)\rightarrow 0$ as $\|\nabla I\|\rightarrow\infty$, and initial condition $I(x,y,0)$ of the PDE is the original image. If $f(\|\nabla I\|):=1$, this becomes Gaussian filtering \citep{koenderink84,hummel87}. The imaginary time $t$ is for the diffusion of pixel values of a single image. Many possible choices exist for $f(\cdot)$, such as $f(x)=1\left/\left(1+\frac{x^2}{k^2}\right)\right.$ or $f(x)=\exp\left\{\frac{-x^2}{k^2}\right\}$, where these edge-stopping functions are the same at all imaginary times $t$. The quantity $k$ is analogous to a heat conduction coefficient or smoothing parameter $h$. The PDE in Equation (\ref{eq:fullpde}) has no analytical solution for most initial and boundary conditions, but the behaviour of the system is approximated by a numerical solution.

Kernel smoothing methods for change-point data are used in a variety of contexts, and the employment of these methods is predicated on both the target regression model and definition of a change-point for that model \citep{gijbels07}. As the dimensions of the regressors increases, the scope of definitions for a change-point broadens; a one-dimensional change-point is intuitively straightforward, where a five-dimensional change-point is hard to conceptualize. We use the generalized nonparametric definition of the change-point regression model $Y_i = g(X_i)+\epsilon_i$ where we have a set of multivariate $q$-explanatory and univariate response continuous data $\left\{(X_i,Y_i)\right\}_{i=1}^n$, and $g$ is a smooth nonlinear regression function with unknown form and unknown set of abrupt changes in the regression function $D$ such that $\mathbb{P}(D)=0$.

One kernel smoothing method for change-point data under this regression model is jump regression \citep{qiu05}. Jump regression estimates the relationship between variables by first detecting jump locations and then smoothing regression functions \citep{qiu91} and derivatives \citep{qiu98}. Jump regression methods fall into two categories: (1) algorithms that detect jumps, define sub-intervals between jumps, and smooth over those intervals, and (2) algorithms that detect jumps, determine the magnitude of the jumps, subtract that magnitude to remove change-points in the data and smooth over the resulting function \cite{qiu91}. Jump regression hinges on the ability to identify change-points before smoothing. For extensive detail on jump regression analysis methods, see \cite{qiu05} and references therein.

Another method for regression estimation of change-point data is the split linear fit that combines left, right, and center linear fits to make smooth functions \citep{mcdonald86,hall92}. This work leads to the study of one-sided kernel functions with asymmetric support for change-point regression, and extends to change-points in the derivative of the regression function \citep{muller92}. Change-point locations are also estimated from the derivative of the local constant estimator \citep{gijbels99} with appropriate bandwidth selection \citep{gijbels04}. The jump regression method above can be extended to local kernel smoothing methods by extending the local linear minimization problem to a piece-wise minimization problem \citep{fan96,qiu03}. The piecewise linear minimization method is used for surface estimation with known or unknown number of jumps locations \citep{qiu07}.

A more robust set of methods that implicitly detect change-points while estimating a regression function is a framework of estimators that consists of $M$-estimators, local $M$-smoothers, Bayesian / regularization / diffusion filtering, and bilateral filtering \citep{mrazek06}. These methods are typically used for for image processing applications, particularly in the area of preserving boundary conditions. The local constant kernel estimator--``adaptive smoothing" in the image processing literature--is similar to anisotropic diffusion and bilateral filters, and can be viewed as a conceptual bridge between the two filters \citep{barash02}. The advantage of bilateral filtering is that optimized versions exist, which do not require a smoothing parameter selection when using a median filter \citep{weiss06}.

Change-point data analysis from the kernel regression perspective currently has limitations. $M$-estimators as edge preserving estimators have been studied for local constant \citep{chu98}, local linear \citep{rue02,lin09}, and local polynomial \citep{hwang04} regression. The $M$-estimators that detect edges and preserve them during smoothing are used in image processing by computer scientists, but lack a critical component sought by statisticians interested in boundary-preserved smoothing; these estimators do not converge to the data generating process $g$ in probability. In this paper, we introduce estimators inspired by the above framework of estimators, that we have named ``anisotropic" smoothers that we show to converge in probability. Our estimators include a range kernel--a kernel function that incorporates the range of the data-- that allows the estimators to anisotropically smooth in the domain. We investigate using the data generating process $g$ inside the range kernel and demonstrate convergence in probability of that estimator. We define change-points as an abrupt change in the regression function over a set $D\in R^q$ such that $\mathbb{P})D=0$. Anisotropic smoothing has potential for local polynomial estimation of a multivariate regression function with multivariate regressors, using a data-driven method that implicitly detects change-points and naturally smooths change-point regression functions. The anisotropic smoothing methods described herein provide new knowledge and insight into the consistency of estimating a regression function with change-points. The path from imagination to realization of a family of consistent anisotropic kernel estimators for change-point regression functions starts here, where we make the first steps by studying anisotropic local constant kernel estimators for multivariate regressors and a univariate outcome.

\section{Anisotropic smoothing for kernel regression}
Consider the $q$-D multivariate regressors $X_i=(X_{1i},X_{2i},\ldots,X_{qi}),\: q\in\mathbb{N}$, univariate outcome $Y_i,i\in\{1,\ldots,n\}$, and the regression model
\begin{eqnarray} \label{eqn:nonparRegModel}
  Y_i=g(X_i)+\epsilon_i,
\end{eqnarray}
 where $g(\cdot)$ is a nonlinear smooth function. The noise $\epsilon_i$ is assumed to have mean 0, variance $\sigma^2$, and $\textup{Cov}\left(\epsilon_i,\epsilon_j\right)=0,$ for $i\neq j$.  The minimization problem for the local constant kernel estimator is
\begin{eqnarray}\label{eqn:LCminimization}
  \min_a\sum_{j=1}^n(Y_j-a)^2K\left(\frac{X_i-x}{h}\right),
\end{eqnarray}
which can be characterized as a local $M$-smoother minimization problem written as
\begin{eqnarray}\label{eqn:localMSmoother}
  \min_{\{u_1,\ldots,u_n\}}\sum_{i=1}^n\sum_{j\in\mathcal{B}_i}\Psi(u_i-Y_j)w(X_i-X_j),
\end{eqnarray}
Without loss of generality, we assume second order for kernel functions and each satisfies the properties:
\begin{eqnarray*}
  (i) & \sup\limits_{-\infty<x<\infty}|k(x)|< \infty, \int_{-\infty}^\infty|k(x)|dx<\infty, \lim\limits_{x\rightarrow\infty}|xk(x)|dx=0,\\
  (ii) & \int_{-\infty}^\infty k(x)dx=1, \\
  (iii) & k(x)=k(-x) \Rightarrow \int_{-\infty}^\infty xk(x)dx=0.
\end{eqnarray*}
The order of the kernel function can affect bias for the local constant estimator, and in turn rate of convergence. The solution to the minimization problem in (\ref{eqn:LCminimization}) is the local constant estimator
\begin{eqnarray}
\widehat{g}(x)\equiv\widehat{a}(x)=\frac{\sum_{i=1}^nY_iK\left(\frac{X_i-x}{h}\right)}{\sum_{i=1}^nK\left(\frac{X_i-x}{h}\right)}.
\end{eqnarray}
The local constant estimator is a locally weighted average of $Y_i$'s using $\widehat{g}(x)=\sum_{i=1}^nY_iw_i$, with weights $K\left(\frac{X_i-x}{h}\right)/\sum_{i=1}^{n}K\left(\frac{X_i-x}{h}\right)$, to estimate the regression function $g(x)$. However, nonnegative kernels do not exist for orders $\nu>2$ \citep{li07}. Using a second order kernel, the local constant estimator have a consistent rate of convergence to $g(x)$ of $O\left(n^{-2/(q+4)}\right)$ \citep{li07}. The anisotropic nonparametric regression estimator is a result from a modification to the minimization problem in Equation (\ref{eqn:LCminimization}). Consider the minimization problem
\begin{eqnarray}\label{eqn:aniMinimization}
& \min_b\sum_{j=1}^n(Y_j-b)^2K\left(\frac{X_j-x}{h}\right)k\left(\frac{Y_j-y}{h_{q+1}}\right),
\end{eqnarray}
and solution
\begin{eqnarray} \label{eqn:weirdTwoParameters}
  \widehat{b}(x,y)=\frac{\sum_{i=1}^nY_iK\left(\frac{X_i-x}{h}\right)k\left(\frac{Y_i-g(x)}{h_{q+1}}\right)}{\sum_{i=1}^nK\left(\frac{X_i-x}{h}\right)k\left(\frac{Y_i-g(x)}{h_{q+1}}\right)}.
\end{eqnarray}
The kernel function $k\left(\frac{Y_i-y}{h_{q+1}}\right)$ is known in the image processing literature as a range (or tonal) kernel \citep{mrazek06}. Range kernels allow for the range of outcome $Y_i$ to be included in the kernel weighting procedure. Points more local in both the domain and range of the regression function contribute more to the smoothing of the regression function.

Equation (\ref{eqn:aniMinimization}) is almost a local $M$-smoother minimization problem, but not quite. If we set $y:=b$ then this would be an $M$-estimator. This equation is similar to the bilateral filter from image processing, but our goal is to study it from a regression perspective. We want to understand the entire function over all $x$, and not just estimate at the observed data $X=x$. We want the regression function estimate to not depend on $y$, since we are interested in predicting $Y|X=x$ and to develop a consistent change-point regression estimator. We propose replacing $y$ with $g(x)$, the true underlying regression function. Consider rewriting the estimator in (\ref{eqn:weirdTwoParameters}) as
\begin{eqnarray*} \label{eqn:closerToAlc}
  \widehat{b}(x)=\frac{\sum_{i=1}^nY_iK\left(\frac{X_i-x}{h}\right)k\left(\frac{Y_i-g(x)}{h_{q+1}}\right)}{\sum_{i=1}^nK\left(\frac{X_i-x}{h}\right)k\left(\frac{Y_i-g(x)}{h_{q+1}}\right)}.
\end{eqnarray*}
The range kernel no longer depends on our outcome variable $y$, retains the anisotropic filter through the range kernel function, and we write our estimator as a function without $y$. However, the estimator contains $Y_i$ embedded inside the kernel function and the error $\epsilon_i$ for the regression model in Equation (\ref{eqn:nonparRegModel}) becomes recursively embedded in the estimator. To avoid this issue, we propose replacing $Y_i$ with $E[Y_i|X_i]=g(X_i)$. Therefore, the anisotropic local constant estimator is
\begin{eqnarray}\label{eqn:alc}
  \widehat{g}(x)=\frac{\sum_{i=1}^nY_iK\left(\frac{X_i-x}{h}\right)k\left(\frac{g(X_i)-g(x)}{h_{q+1}}\right)}{\sum_{i=1}^nK\left(\frac{X_i-x}{h}\right)k\left(\frac{g(X_i)-g(x)}{h_{q+1}}\right)},
\end{eqnarray}
where it can be viewed as an anisotropic extension of the local constant kernel estimator. The intuition behind the anisotropic local constant estimator is that we are locally averaging the regression function over $x$ {\bf and} the regression function itself $g(x)$. Data that are farther away in $x$ or $g(x)$ from $X_i$ are given lower weights than more local points. If there is a large jump in the data, points local in $x$ but distant in $y$ will contribute less and we are able to smooth across jump regions. This estimator is notably similar to the nonparametric estimator for the conditional cumulative distribution function \citep{li08}, where the conditional cumulative distribution function estimator is considered to be more robust for estimating regression functions, particularly in the presence of censoring \cite[ch. 6]{li07}.

\section{Weak consistency of the anisotropic local constant estimator} \label{sec:rateOfConvergenceALC}
We show the weak consistency of the anisotropic local constant estimator in Equation (\ref{eqn:alc}), under the following assumptions:
\begin{itemize}
 \item[T1] The kernel function is second-order ($\nu=2$). This benchmark for the order of the kernel will allow us to compare rates of convergence with the isotropic kernel estimator.
 \item[T2]  The regressors $X_i$ are fixed and not random. For imagery data over time, it is natural to assume pixel locations are fixed in space. This assumption is typical for kernel estimator consistency derivations in the literature \citep{ruppert94}.
 \item[T3] Bandwidths are of similar orders of magnitude. This ensures that $h_1,h_2,\ldots,h_q\rightarrow 0$ at the same rate. We also assume the bandwidth associated with the range kernel goes to zero ($h_{q+1}\to 0$) at a faster rate than any of $h_1,h_2,\ldots,h_q$.
 \item[T4] The design of $X_i$ is equally spaced by $\frac{1}{n_j}$, suck that $\Delta x = \Delta x_1 \Delta x_2\ldots\Delta x_q=(X_{1i}-X_{1(i-1)})(X_{2(i-1)}-X_{2(i-1)})\ldots(X_{qi}-X_{q(i-1)}) = \frac{1}{n}$.
\end{itemize}
{\bf Theorem 1 }{\it  Under the assumptions T1-T3 above, and the assumptions that $x$ is an interior point and $g(x)$ is at least three times differentiable (except on a set $D$ that forms a set of points (univariate $x$) or curves (multivariate $x$) such that $\mathbb{P}(D)=0$), then as $n\rightarrow\infty$, $h_j\rightarrow0$, $n_j h_j\rightarrow\infty,$ $\forall~ j\in\{1,2,\ldots,q,q+1\}$, we have, for any $\epsilon>0$, }
\begin{eqnarray*}
 \lim_{n\to\infty}\mathbb{P}\left(|\widehat{g}(x)-g(x)|<\epsilon\right)=1,
\end{eqnarray*} 
{\it where $n=n_1\times n_2\times\ldots\times n_q=n_{q+1}$.}

\noindent {\it Proof.} To prove weak consistency, we need to show that
\begin{eqnarray*}
   \lim_{n\to\infty}\textup{Var}(\widehat{g}(x))=0,\\
   \lim_{n\to\infty}E[\widehat{g}(x)-g(x)]=0.
\end{eqnarray*}
First, consider rewriting the estimator in (\ref{eqn:alc}) as 
\begin{eqnarray*}
  \widehat{g}(x)=\frac{\frac{1}{n}\sum_{i=1}^nY_iK\left(\frac{X_i-x}{h}\right)k\left(\frac{g(X_i)-g(x)}{h_{q+1}}\right)}{\frac{1}{n}\sum_{i=1}^nK\left(\frac{X_i-x}{h}\right)k\left(\frac{g(X_i)-g(x)}{h_{q+1}}\right)},
\end{eqnarray*}
where $Y_i$ is univariate and $X_i$ is $(q\times 1)$ vector. Substituting that $Y_i=g(X_i)+\epsilon_i$ yields
\begin{eqnarray*}
  \widehat{g}(x)=\frac{\frac{1}{n}\sum_{i=1}^n(g(X_i)+\epsilon_i)K\left(\frac{X_i-x}{h}\right)k\left(\frac{g(X_i)-g(x)}{h_{q+1}}\right)}{\frac{1}{n}\sum_{i=1}^nK\left(\frac{X_i-x}{h}\right)k\left(\frac{g(X_i)-g(x)}{h_{q+1}}\right)}.
\end{eqnarray*}
Taking the expectation of the above equation--conditional on the $X_i$'s and the assumption that $E[\epsilon_i]=0$--gives
\begin{eqnarray*}
  E\left[\widehat{g}(x)\right]&=&\frac{\frac{1}{n}\sum_{i=1}^ng(X_i)K\left(\frac{X_i-x}{h}\right)k\left(\frac{g(X_i)-g(x)}{h_{q+1}}\right)}{\frac{1}{n}\sum_{i=1}^nK\left(\frac{X_i-x}{h}\right)k\left(\frac{g(X_i)-g(x)}{h_{q+1}}\right)}.
\end{eqnarray*}
Under assumption T4 and as each $n_i$ becomes large, we approximate these Riemann sums as integrals in  
\begin{eqnarray*}
E\left[\widehat{g}(x)\right]&\approx&\frac{\int g(w)K\left(\frac{w-x}{h}\right)k\left(\frac{g(w)-g(x)}{h_{q+1}}\right)dw}{\int K\left(\frac{w-x}{h}\right)k\left(\frac{g(w)-g(x)}{h_{q+1}}\right)dw},
\end{eqnarray*}
where $w = (w_1,w_2,\ldots,w_q)$, $dw=dw_1dw_2\ldots dw_q$. Using the substitution $z_i=\frac{w_i-x_i}{h_i}$, $dz_i=\frac{1}{h_i}dw_i$, $z=(z_1,z_2,\ldots,z_q)$, $dz=dz_1dz_2\ldots dz_q$, $hdz=h_1dz_1h_2dz_2\ldots h_qdz_q$ and $zh=(z_1h_1,z_2h_2,\ldots,z_qh_q)$ gives
\begin{eqnarray} \label{eqn:expectation1}
  E\left[\widehat{g}(x)\right]&\approx&\frac{\int g(zh+x)K(z)k\left(\frac{g(zh+x)-g(x)}{h_{q+1}}\right)hdz}{\int K(z)k\left(\frac{g(zh+x)-g(x)}{h_{q+1}}\right)hdz}.
\end{eqnarray}
A Taylor series expansion of $g(zh+x)$ about $z=0$ is 
\begin{eqnarray} \label{eqn:taylorSeries1}
g(zh+x)=g(x)+zh\cdot\nabla g(x)+\frac{1}{2}(zh)' Hzh+\ldots,
\end{eqnarray}
where
\begin{eqnarray*}
g_{z_i}(x)=\left.\frac{\partial}{\partial z_i} g(z)\right|_{z=x},\: g_{z_iz_j}(x)=\left.\frac{\partial^2}{\partial z_i\partial z_j} g(z)\right|_{z=x},\\
\nabla g(x) = (g_{z_1}(x),g_{z_2}(x),\ldots,g_{z_q}(x)),\\
zh\cdot\nabla g(x)=\sum_{j=1}^qz_jh_jg_{z_j}(x) \propto O(h_\star),\\
H = (H_{ij})_{q\times q}, H_{ij}=g_{z_iz_j}(x),\\
(zh)' Hzh = \sum_{j=1}^q\sum_{k=1,k\neq j}^q 2g_{z_jz_k}(x)z_jz_kh_jh_k+\sum_{j=1}^q\ g_{z_jz_j}(x)z_j^2h_j^2 \propto O(h_\star^2),
\end{eqnarray*}
and we consider $h_\star$ to be any bandwidth $h_j,$ $j\in \{1,\dots,q\}$. Inserting the Taylor series expansion in Equation (\ref{eqn:taylorSeries1}) into Equation (\ref{eqn:expectation1}) yields
\begin{eqnarray}
E\left[\widehat{g}(x)\right]&\approx&\frac{\int \left(g(x)+zh\cdot\nabla g(x)+\frac{1}{2}(zh)' Hzh+\ldots\right)K(z)k\left(\frac{g(zh+x)-g(x)}{h_{q+1}}\right)dz}{\int K(z)k\left(\frac{g(zh+x)-g(x)}{h_{q+1}}\right)dz}, \nonumber \\
&=& g(x)+ \sum_{i=1}^q h_ig_{z_i}(x)\frac{\int K(z)k\left(\frac{g(zh+x)-g(x)}{h_{q+1}}\right)zdz}{\int K(z)k\left(\frac{g(zh+x)-g(x)}{h_{q+1}}\right)dz} + O(h_i^2).\label{eqn:alcBias}
\end{eqnarray}
Recall the assumptions for the kernel function $k$ that as $w\rightarrow\pm\infty$, $k(w)\rightarrow 0$,and $k(w)<\infty$ for $w \in \mathbb{R}$. Consider the Taylor series expansion inside the kernel function 
\begin{eqnarray*} \label{eqn:oneDkerneltaylor}
k\left(\frac{g(zh+x)-g(x)}{h_{q+1}}\right)=k\left(\frac{zh\cdot\nabla g(x)+\frac{1}{2}(zh)' Hzh+O(h^3)}{h_{q+1}}\right).
\end{eqnarray*}
We denote the direction where there is a jump in $g(x)$ as $\star$. Specifically at a jump in $g(x)$, we have $g_{x_\star}(x)$ is very large such that $g_{x_\star}(x)\rightarrow\pm\infty$. We also have that as $n\to\infty$, $h_\star\rightarrow0$. Either $g_{x_\star}(x)h_\star\rightarrow \pm\infty$ or $g_{x_\star}(x)h_\star\rightarrow 0$ but slower than $h_{q+1}$, thus the first term $\frac{g_{x_\star}(x)h_i}{h_{q+1}}\rightarrow\pm\infty$. Similarly, $g_{x_{\star\star}}(x)\rightarrow\pm\infty$ gives that the second term either goes to 0 or $\pm\infty$. Thus, we have that $k\left(\frac{g(zh+x)-g(x)}{h_{q+1}}\right)\rightarrow0$ for $h_\star\rightarrow0$ and $g_{x_\star}(x)\rightarrow0$. The other terms are order $O\left(\frac{h_\star^2}{h_{q+1}}\right)=O(h_\star)\rightarrow 0$. We assume that as $n\rightarrow\infty$, then $h_\star \rightarrow0$. Therefore as $n\rightarrow\infty$, we get that $E\left[\widehat{g}(x)\right]=g(x)$ and that the anisotropic local constant estimator is asymptotically unbiased. 
 
The variance of the estimator is
\begin{eqnarray*}
\var\left(\widehat{g}(x)\right) &=& \var\left(\frac{\sum_{i=1}^nY_iK\left(\frac{X_i-x}{h}\right)k\left(\frac{g(X_i)-g(x)}{h_{q+1}}\right)}{\sum_{i=1}^nK\left(\frac{X_i-x}{h}\right)k\left(\frac{g(X_i)-g(x)}{h_{q+1}}\right)}\right), \\ 
 &=& \var\left(\frac{\sum_{i=1}^n(g(X_i)+\epsilon_i)K\left(\frac{X_i-x}{h}\right)k\left(\frac{g(X_i)-g(x)}{h_{q+1}}\right)}{\sum_{i=1}^nK\left(\frac{X_i-x}{h}\right)k\left(\frac{g(X_i)-g(x)}{h_{q+1}}\right)}\right).
\end{eqnarray*}
Since $X_i$ is fixed and $\var(\epsilon_i)=\sigma^2$, this simplifies to
\begin{eqnarray*}
\var\left(\widehat{g}(x)\right) &=&\sigma^2\frac{\sum_{i=1}^n\left[K\left(\frac{X_i-x}{h}\right)k\left(\frac{g(X_i)-g(x)}{h_{q+1}}\right)\right]^2}{\left[\sum_{i=1}^nK\left(\frac{X_i-x}{h}\right)k\left(\frac{g(X_i)-g(x)}{h_{q+1}}\right)\right]^2}, \\
&=& \frac{\sigma^2}{n}\frac{\frac{1}{n}\sum_{i=1}^n\left[K\left(\frac{X_i-x}{h}\right)k\left(\frac{g(X_i)-g(x)}{h_{q+1}}\right)\right]^2}{\left[\frac{1}{n}\sum_{i=1}^nK\left(\frac{X_i-x}{h}\right)k\left(\frac{g(X_i)-g(x)}{h_{q+1}}\right)\right]^2}. 
\end{eqnarray*}
Taking $n$ to be very large allows us to approximate these sums as integrals in
\begin{eqnarray*}
  \var\left(\widehat{g}(x)\right) &\approx& \frac{\sigma^2}{n}\frac{\int\left[K\left(\frac{w-x}{h}\right)k\left(\frac{g(w)-g(x)}{h_{q+1}}\right)\right]^2dw}{\left[\int K\left(\frac{w-x}{h}\right)k\left(\frac{g(w)-g(x)}{h_{q+1}}\right)dw\right]^2}.
\end{eqnarray*}
Substituting $z=\frac{w-x}{h}$ as before yields
\begin{eqnarray} 
  \var\left(\widehat{g}(x)\right) &\approx& \frac{\sigma^2}{n}\frac{\int\left[K(z)k\left(\frac{g(zh+x)-g(x)}{h_{q+1}}\right)\right]^2hdz}{\left[\int K(z)k\left(\frac{g(zh+x)-g(x)}{h_{q+1}}\right)hdz\right]^2},\nonumber\\
  &=& \frac{\sigma^2}{nh}\frac{\int\left[K(z)k\left(\frac{g(zh+x)-g(x)}{h_{q+1}}\right)\right]^2dz}{\left[\int K(z)k\left(\frac{g(zh+x)-g(x)}{h_{q+1}}\right)dz\right]^2}.\label{eqn:alcVariance}
\end{eqnarray}
We have assumed that as $n\rightarrow0$ and $h\rightarrow0$, then $nh\rightarrow\infty$. This assumption implies that $\var\left(\widehat{g}(x)\right)\rightarrow 0$ as $n\rightarrow\infty$. Therefore, $\widehat{g}(x)$ is a consistent estimator of $g(x)$.

\hfill\ensuremath{\blacksquare}

We have established that the anisotropic local constant estimator in Equation (\ref{eqn:alc}) converges in probability to the data generating process $g(x)$ under certain conditions. The next step is to establish the rate that the anisotropic local constant estimator converges to the underlying data generating process compared to competing methods. 
\subsubsection{Rate of convergence of the anisotropic local constant estimator}
The rate of convergence is important when comparing estimation methods, as it guides a practitioner on which estimator has the minimum expected squared error for a given sample, assuming the regression model is correctly specified. Typically, The fewer assumptions on the structure of the data generating process the slower an estimator converges. A pointwise estimate of the mean squared error (MSE) is used to find the optimal rate of convergence. Using the expressions in Equations (\ref{eqn:alcBias}) and (\ref{eqn:alcVariance}) for the bias and variance, the MSE for this estimator is proportional to the sample size and lowest order of bandwidth by
\begin{eqnarray}\label{eqn:alcMSE}
  \textup{MSE}(\widehat{g}(x)) =\textup{bias}\left(\widehat{g}(x)\right)^2 + \var\left(\widehat{g}(x)\right) \propto h_\star ^2+ \frac{1}{nh_\star ^{q}}.
\end{eqnarray}
The assumption that each bandwidth of the kernel functions has the same magnitude and converges to zero at the same rate allows us to treat each $h_1,\ldots,h_q$ as the ``same" $h_\star $ in this context. Since the bandwidth is a smoothing parameter we specify, we calculate an optimal smoothing parameter that minimizes the MSE and determine the relationship of sample size to the optimal smoothing parameter. Taking the first derivative of the MSE with respect to bandwidth $h_\star $ yields
\begin{eqnarray}
  & \frac{\partial MSE}{\partial h_\star }\propto 2h_\star +\frac{-1}{nh_\star ^{q+1}},\nonumber\\
  \Rightarrow & 0 \propto 2h_\star +\frac{-1}{nh_\star ^{q+1}},\nonumber\\
  \Rightarrow & h_\star \propto n^{-\frac{1}{q+2}}.\label{eqn:optiband}
\end{eqnarray}
This shows that as $n\rightarrow\infty$, we require the bandwidth $h_i\rightarrow 0$, which we assumed for the proof of Theorem 1. Inserting the optimal bandwidth in Equation (\ref{eqn:optiband}) into (\ref{eqn:alcMSE}) shows the MSE converges to zero at a rate $O\left(n^{\frac{-2}{(q+2)}}\right)$. This implies the rate of convergence for our estimator is $O\left(n^{\frac{-1}{(q+2)}}\right)$, which is slower than the isotropic estimator's rate $O\left(n^{\frac{-2}{(q+4)}}\right)$.

Recall that the parametric rate of convergence is $O\left(n^{-\frac{1}{2}}\right)$, so the parametric model should converge to the data generating process faster than the isotropic estimator. However, this is for a {\bf correctly} specified parametric model. If the parametric model is misspecified, then the parametric estimator never converges to the data generating process. In a similar way, the isotropic estimator should outperform the anisotropic, as long as it is the correctly specified estimator. If there are change-points in the regression function, we argue that the isotropic estimator is a misspecification and the anisotropic estimator is the correct specification and will outperform the isotropic estimator. 

We are using the data generating process $g(x)$ in a kernel function to estimate itself. This implies we have access to the full knowledge of $g(x)$ that we are trying to estimate. In fact, we do have access to information about $g(x)$ through $Y_i$ and we obtain that information through a pilot estimator $\widehat{g}(x)$.  
\subsection{Pilot estimators for the anisotropic local constant estimator} \label{sec:pilotEstim}
We have shown the weak consistency of the $q$-dimensional anisotropic local constant estimator when using the true regression function $g(x)$. This estimator allows local information to be filtered if there is a large change-point in the regression function $g(x)$. However, in practice, we do not have access to the full information of $g(x)$. In order to use this estimator, we propose using a pilot estimator $\tilde{g}(x)$. The anisotropic local constant estimator becomes
\begin{eqnarray}\label{eqn:realALC}
  \widehat{g}(x)=\frac{\sum_{i=1}^nY_iK\left(\frac{X_i-x}{h}\right)k\left(\frac{\tilde{g}(X_i)-\tilde{g}(x)}{h_{q+1}}\right)}{\sum_{i=1}^nK\left(\frac{X_i-x}{h}\right)k\left(\frac{\tilde{g}(X_i)-\tilde{g}(x)}{h_{q+1}}\right)}.
\end{eqnarray}
Equation (\ref{eqn:realALC}) is no longer the same estimator as Equation (\ref{eqn:alc}), so we need to re-establish weak consistency. We show the anisotropic local constant estimator that uses $\tilde{g}(x)$ as a pilot estimator also converges in probability to $g(x)$ under the assumption that the pilot estimator has the property $\tilde{g}(x)\overset{P}{\to}g(x)$. 

\noindent {\bf Theorem 2 } {\it Assume that $\tilde{g}(x)\overset{P}{\to}g(x)$. For the estimator in (\ref{eqn:realALC}), then $\widehat{g}(x)\overset{P}{\to}g(x)$}

\noindent {\it Proof.} Consider the estimator in Equation (\ref{eqn:realALC}) with a Taylor series expansion from Equation (\ref{eqn:taylorSeries1}) for $\tilde{g}(X_i)$ about $x$
\begin{eqnarray} \label{eqn:pilotproof1}
\widehat{g}(x,y)=\frac{\sum_{i=1}^nY_iK\left(\frac{X_i-x}{h}\right)k\left(\frac{(X_i-x)\cdot\nabla \tilde{g}(x)+\frac{1}{2}(X_i-x)' \tilde{H}(X_i-x)+\tilde{R}(x)}{h_{q+1}}\right)}{\sum_{i=1}^nK\left(\frac{X_i-x}{h}\right)k\left(\frac{(X_i-x)\cdot\nabla \tilde{g}(x)+\frac{1}{2}(X_i-x)' \tilde{H}(X_i-x)+\tilde{R}(x)}{h_{q+1}}\right)}.
\end{eqnarray}
The Continuous Mapping Theorem \cite{grimmett01} yields 
\footnotesize\begin{eqnarray*}
\frac{(X_i-x)\cdot\nabla \tilde{g}(x)+\frac{1}{2}(X_i-x)' \tilde{H}(X_i-x)+\tilde{R}(x)}{h_{q+1}} \overset{P}{\to}\frac{(X_i-x)\cdot\nabla g(x)+\frac{1}{2}(X_i-x)' H(X_i-x)+R(x)}{h_{q+1}},
\end{eqnarray*}
\normalsize and the right side is constant if each term converges, where $\nabla g(x),H,$ and $R(x)$ are bounded. However for this estimator, we assume that as $n\to\infty$ that $h\to 0$. So, we need that $\tilde{g}(x)\to g(x)$ faster than $h\to 0$. Thus, the numerator in Equation (\ref{eqn:pilotproof1}) converges to the random variable numerator in Equation (\ref{eqn:alc}) and the denominator in Equation (\ref{eqn:pilotproof1}) converges to the constant denominator in Equation (\ref{eqn:alc}). Slutsky's Theorem gives that Equation (\ref{eqn:realALC}) converges in probability to Equation (\ref{eqn:alc}) and we are done.

\hfill\ensuremath{\blacksquare}

A candidate pilot estimator that converges in probability is the isotropic local constant kernel estimator. Theorem 2 applies to this pilot estimator, but only with specific assumptions on the bandwidths of the pilot estimator in relation to the bandwidths of the anisotropic estimator. We need the bandwidths for the pilot estimator to converge to zero faster than the bandwidths for the anisotropic estimator, such that $\tilde{g}(x)\to g(x)$ faster than the bandwidths for the anisotropic estimator go to zero. This implies that the anisotropic estimator can be viewed as a corrective method that updates the isotropic estimator if we think the data has change-points. 
\subsection{Iterative estimation}\label{sec:recursiveEstim}
We have shown that the anisotropic local constant converges in probability to the data generating process $g(x)$ if the pilot estimator $\tilde{g}(x)$ converges in probability to $g(x)$. Therefore, we need a pilot estimator that converges to the data generating process and we choose the anisotropic local constant estimator. The algorithm for iterative estimation is:
\begin{enumerate}
  \item Fit the isotropic model using 
  \begin{eqnarray*} \tilde{g}(x)=\frac{\sum_{i=1}^nY_iK\left(\frac{X_i-x}{h}\right)}{\sum_{i=1}^nK\left(\frac{X_i-x}{h}\right)}. \end{eqnarray*}
  \item Fit the anisotropic model using 
  \begin{eqnarray*} \widehat{g}_1(x)=\frac{\sum_{i=1}^nY_iK\left(\frac{X_i-x}{h}\right)k\left(\frac{\tilde{g}(X_i)-\tilde{g}(x)}{h_{q+1}}\right)}{\sum_{i=1}^nK\left(\frac{X_i-x}{h}\right)k\left(\frac{\tilde{g}(X_i)-\tilde{g}(x)}{h_{q+1}}\right)}.\end{eqnarray*}
  \item Iteratively fit the anisotropic model $d$ times using 
  \begin{eqnarray*} \widehat{g}_j(x)=\frac{\sum_{i=1}^nY_iK\left(\frac{X_i-x}{h}\right)k\left(\frac{\widehat{g}_{j-1}(X_i)-\widehat{g}_{j-1}(x)}{h_{q+1}}\right)}{\sum_{i=1}^nK\left(\frac{X_i-x}{h}\right)k\left(\frac{\widehat{g}_{j-1}(X_i)-\widehat{g}_{j-1}(x)}{h_{q+1}}\right)},~ j\in\{2,3,\ldots,d\}.
  \end{eqnarray*}
\end{enumerate}

\section{Simulations}
\subsection{One-dimension}
\begin{figure}[h!]

{\centering \includegraphics[width=1\textwidth]{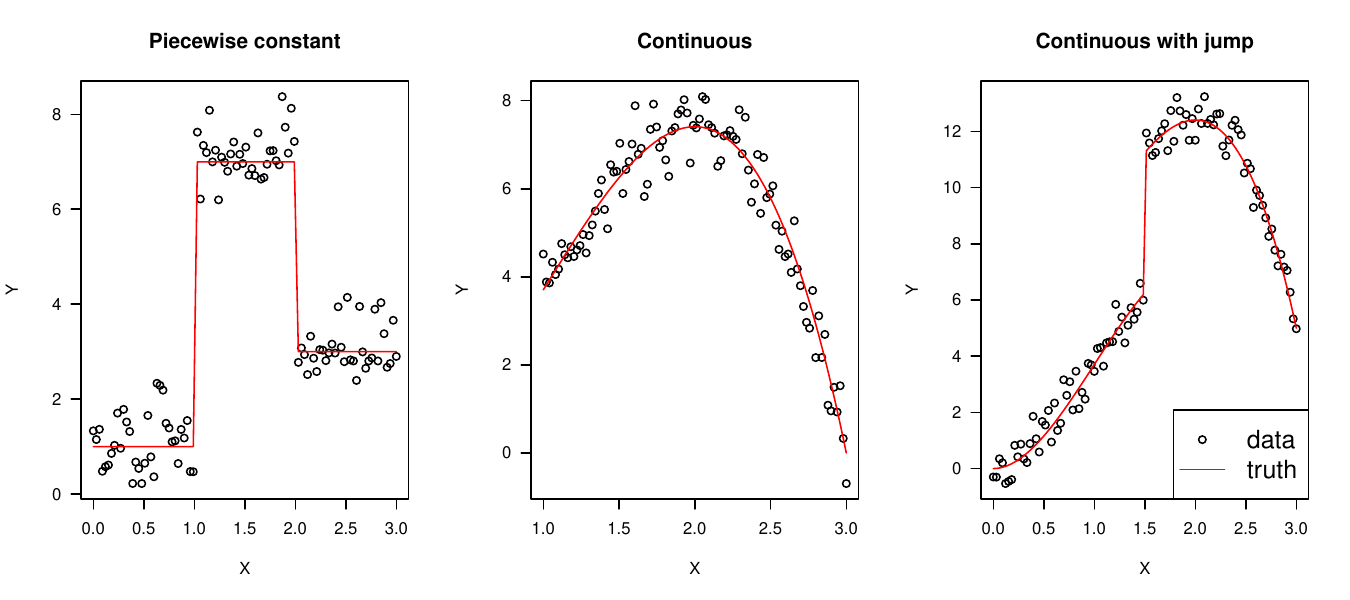} 

}

\caption[The three one-dimensional data generating processes used for simulations$.~$The left most figure is the piecewise constant regression function$.~$The middle figure is the continuous regression function$.~$The right most figure is the same continuous regression function with a jump at $X=1.5.$]{The three one-dimensional data generating processes used for simulations$.~$The left most figure is the piecewise constant regression function$.~$The middle figure is the continuous regression function$.~$The right most figure is the same continuous regression function with a jump at $X=1.5.$}\label{fig:dataTypes}
\end{figure}

We have chosen a few simple simulated change-point regression functions that allow us to explore the performance of our estimator on change-point data. The simulated data use a piecewise constant function with two change-points and a continuous function with and without change-points. The three data generating processes used in the simulation are shown in Figure \ref{fig:dataTypes} using the functions
\begin{eqnarray*} \label{eqn:dataGeneratingProcesses}
\textup{Piecewise constant:} & g(x) = \left\{\begin{matrix} 1, & 0\leq x\leq 1, \\ 7, & 1 < x\leq 2, \\ 3, & 2<x\leq 3, \end{matrix}\right. \nonumber\\
\textup{Continuous:} & g(x) = 50\left[\left(\frac{x}{3}\right)^2-\left(\frac{x}{3}\right)^3\right],~ 0\leq x\leq 3,\\
\textup{Continuous with jump:} & g(x) =\left\{\begin{matrix}50\left[\left(\frac{x}{3}\right)^2-\left(\frac{x}{3}\right)^3\right], &0\leq x\leq 1.5,\\50\left[\left(\frac{x}{3}\right)^2-\left(\frac{x}{3}\right)^3\right],& 1.5< x\leq 3,\end{matrix}\right.\nonumber
\end{eqnarray*}
and i.i.d. noise $\epsilon\sim N(0,\sigma^2)$. In our Monte Carlo simulations, we consider 125 Monte Carlo replicates with $\sigma\in\{0.1,0.5,1,2\}$. The Monte Carlo simulation algorithm is:
\begin{enumerate}
  \item Simulate $n$ $N(0,\sigma^2)$ random variates $\epsilon_i$ and calculate $Y_i=g(X_i)+\epsilon_i$ such that $X_i$ start at 0 and are evenly spaced in the interval $0\leq X_i\leq 3$.
  \item Fit the isotropic model using 
  \begin{eqnarray*} \tilde{g}(x)=\frac{\sum_{i=1}^nY_iK\left(\frac{X_i-x}{h}\right)}{\sum_{i=1}^nK\left(\frac{X_i-x}{h}\right)}. \end{eqnarray*}
  \item Fit the anisotropic model using 
  \begin{eqnarray*} \widehat{g}(x)=\frac{\sum_{i=1}^nY_iK\left(\frac{X_i-x}{h}\right)k\left(\frac{\tilde{g}(X_i)-\tilde{g}(x)}{h_{q+1}}\right)}{\sum_{i=1}^nK\left(\frac{X_i-x}{h}\right)k\left(\frac{\tilde{g}(X_i)-\tilde{g}(x)}{h_{q+1}}\right)}.\end{eqnarray*}
  \item Repeat steps 1-3 $M$ times.
\end{enumerate}

The piecewise constant data type simulates a simple change-point regression function that is often analyzed in the image processing literature \cite{chu98,polzehl00,gijbels06}. The continuous regression function without the change-point allows us to investigate the scenario when a change-point estimator is inappropriate, and the continuous regression function with a change-point allows for a more complex regression function on either side of the change-point.

\begin{figure}[h!]

{\centering \includegraphics[width=.8\textwidth]{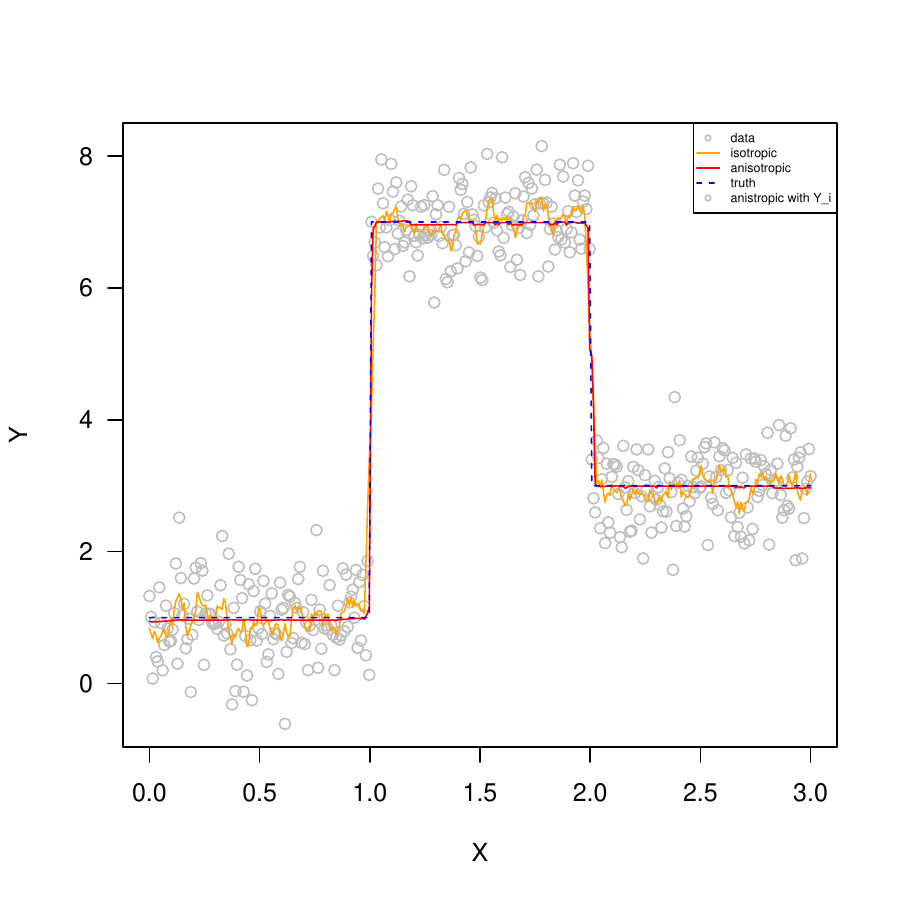} 

}

\caption[Fitting the isotropic and anisotropic local constant estimators to the piecewise constant regression function data and comparing to the underlying data generating process (truth)$.~$This plot uses simulation parameters $n=400$ and $\sigma=0.5$, the uniform kernel function for smoothing, and least-squares cross-validation bandwidth selection method$.$]{Fitting the isotropic and anisotropic local constant estimators to the piecewise constant regression function data and comparing to the underlying data generating process (truth)$.~$This plot uses simulation parameters $n=400$ and $\sigma=0.5$, the uniform kernel function for smoothing, and least-squares cross-validation bandwidth selection method$.$}\label{fig:simulationPlots}
\end{figure}

An example of the isotropic and anisotropic local constant estimators is shown in Figure \ref{fig:simulationPlots}. To determine the theoretical performance of regression estimators, we have looked at the pointwise convergence rate--that is, the rate at which the mean square error between our estimator and the underlying data generating process. In practice, to compare performance of nonparametric estimators on simulated data (since we need to know the function $g$ at all $X_i$) we calculate the mean of the estimated squared errors (MESE) \citep{thompson14} given by
\begin{eqnarray}\label{eqn:MESE}
  \textup{MESE}(\widehat{g})=\frac{1}{n}\sum_{i=1}^n(g(X_i)-\widehat{g}(X_i))^2.
\end{eqnarray}
Three estimators are fit to the simulated data: the isotropic local-constant estimator (LC), our anisotropic local constant estimator with LC as the pilot estimator (ALC), and our anisotropic local constant estimator with the data generating process (the ``truth") as the pilot estimator (ALCT). All three simulated data types are used with four error standard deviations ($\sigma\in\{0.1,0.5,1,2\}$) with $M=125$ Monte Carlo replicates, the uniform kernel, and sample sizes of 400 to 1600. The bandwidth selection method is an AIC cross-validation method \citep{hurvich98}. The relative performance of these estimators is shown in Tables \ref{tbl:simulationResultsTableSigmaPoint1}, \ref{tbl:simulationResultsTableSigmaPoint5}, and \ref{tbl:simulationResultsTableSigma1} using the mean of the MESEs, where the MESE of each Monte Carlo replicate is calculated using Equation (\ref{eqn:MESE}) and then averaged over $M$ Monte Carlo simulations. Tables \ref{tbl:simulationResultsTableSigmaPoint1SD}, \ref{tbl:simulationResultsTableSigmaPoint5SD}, \ref{tbl:simulationResultsTableSigma1SD} show the sample standard deviations of the MESEs. 

\begin{figure}
\includegraphics[width=1\textwidth]{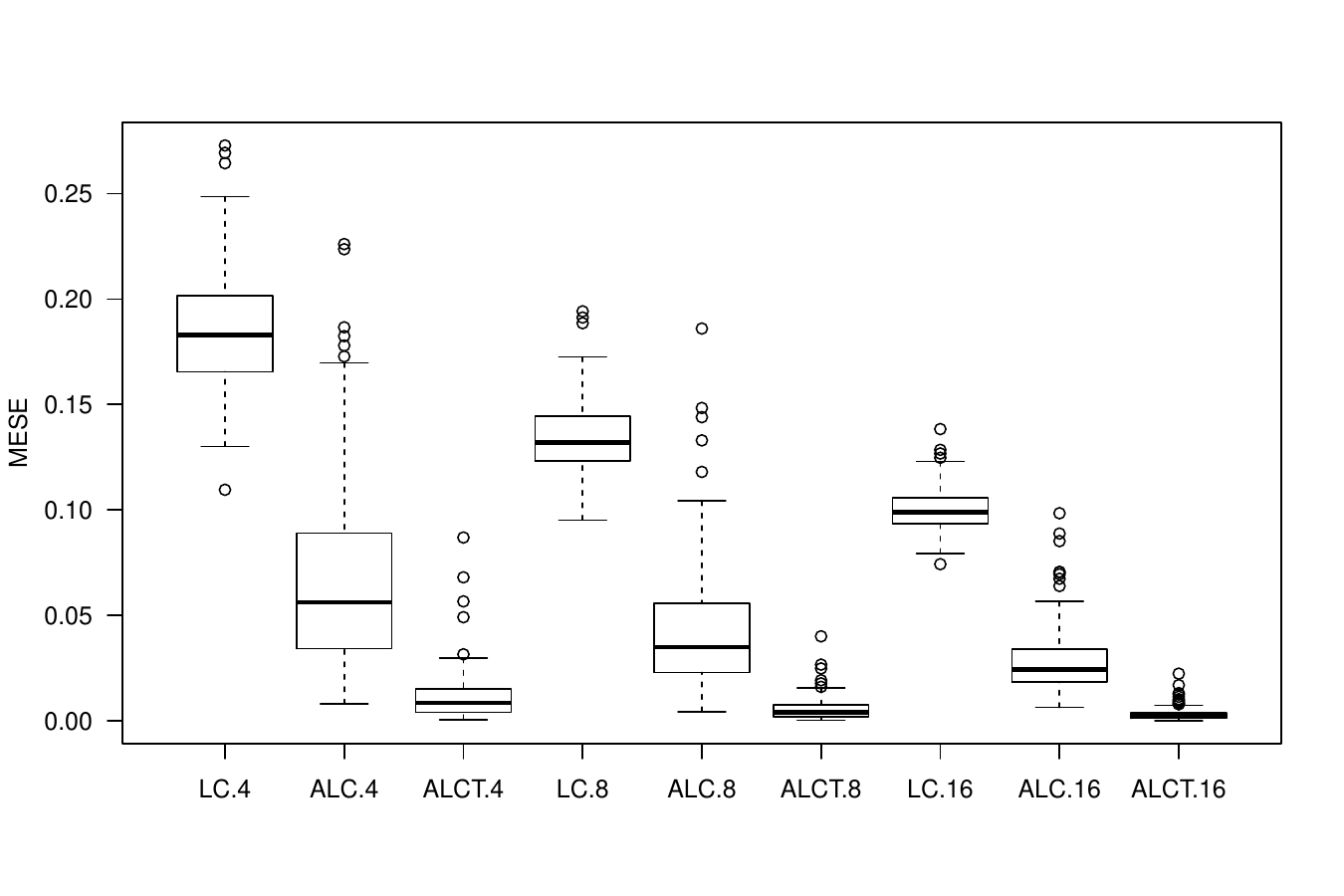} \caption[The MESEs for the three kernel estimators on the piecewise constant function$.~$Note that $.Z$ means $n=Z00$, e.g$.~$LC.4 is the isotropic local constant estimator with $n=400$ and ALC.16 is the anisotropic estimator with $n=1600.$]{The MESEs for the three kernel estimators on the piecewise constant function$.~$Note that $.Z$ means $n=Z00$, e.g$.~$LC.4 is the isotropic local constant estimator with $n=400$ and ALC.16 is the anisotropic estimator with $n=1600.$}\label{fig:boxplotUJ}
\end{figure}

For the piecewise constant regression function data in Table \ref{tbl:simulationResultsTableSigmaPoint1}, the ALC estimator consistently has a smaller MESE than the LC estimator, showing that the ALC estimator improves the fit for the LC estimator by better accounting for the change-point. The ALCT estimator consistently has a smaller MESE than the ALC estimator, showing that anisotropic local constant estimator is further improved by using a better pilot estimator. Given a good pilot estimator, we get relatively better estimates of the data generating process, occluded only by the noise of the data. Table \ref{tbl:simulationResultsTableSigmaPoint1SD} shows that as noise increases, the ALC estimator tends to have a large sample standard deviation for the MESE. However, Figure \ref{fig:boxplotUJ} shows that the ALC is still an overall improvement over the LC estimator.

Table \ref{tbl:simulationResultsTableSigmaPoint5} shows that the ALC estimator has a larger MESE (generally worse fit) than the LC estimator on continuous data that does not contain a jump in the regression function. This comparison demonstrates the effect on smoothing when the ALC is used inappropriately. Interestingly, the ALCT estimator shows an improvement over the LC estimator. This is consistent with similar range kernel methods used to improve regression estimates in the literature \citep{li07}. However, a practitioner may choose not use the ALC estimator in this context, given the LC estimator is performing adequately at estimating the continuous regression function, and change-points are not present in the data. Table \ref{tbl:simulationResultsTableSigmaPoint5SD} shows the sample standard deviations of the MESE for each estimator on the continuous regression function data.
\begin{figure}
\includegraphics[width=1\textwidth]{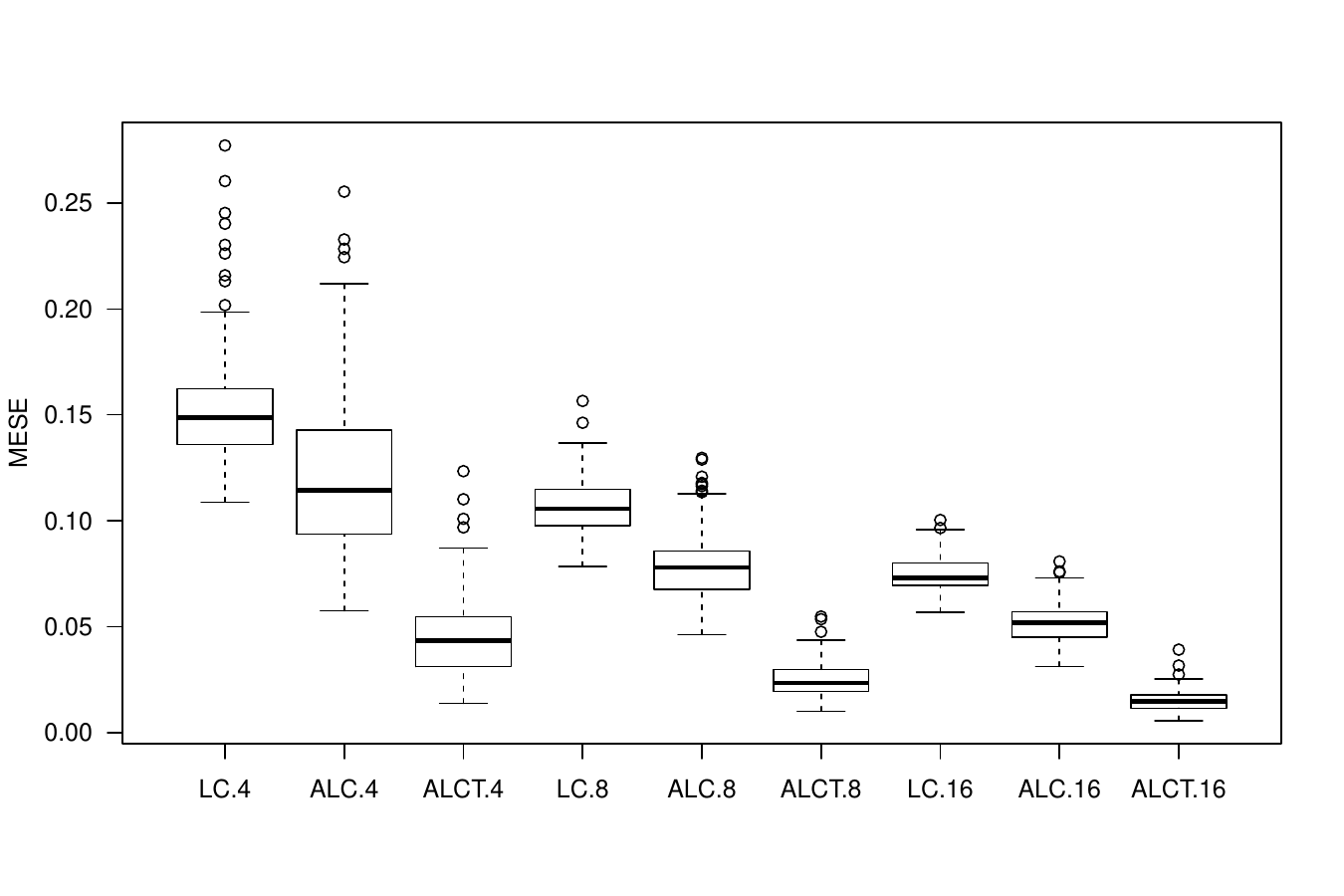} \caption[The MESEs for the three kernel estimators on data from the continuous regression function with a jump$.~$Note that $.Z$ means $n=Z00$, e.g$.~$LC.4 is the isotropic local constant estimator with $n=400$ and ALC.16 is the anisotropic estimator with $n=1600.$]{The MESEs for the three kernel estimators on data from the continuous regression function with a jump$.~$Note that $.Z$ means $n=Z00$, e.g$.~$LC.4 is the isotropic local constant estimator with $n=400$ and ALC.16 is the anisotropic estimator with $n=1600.$}\label{fig:boxplotCJ}
\end{figure}

\begin{figure}[h!]

{\centering \includegraphics[width=.8\textwidth]{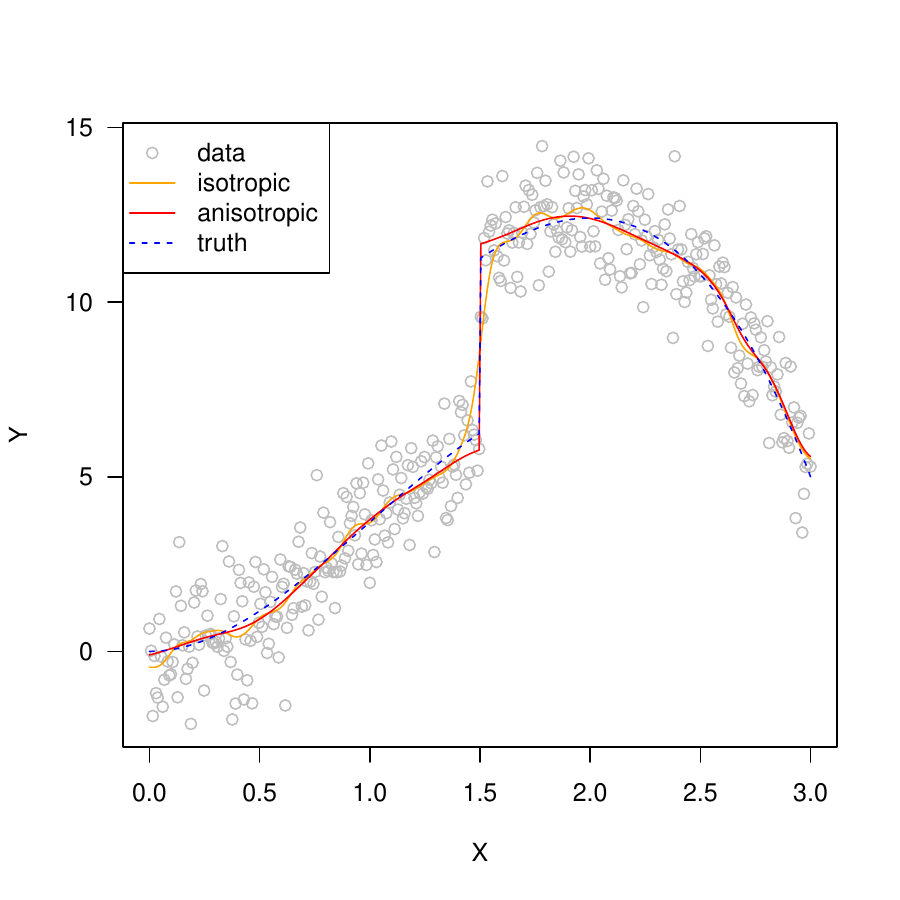} 

}

\caption[Fitting the isotropic and anisotropic local constant estimators to the continuous regression function with a jump data and comparing to the underlying data generating process (truth)$.~$This plot uses simulation parameters $n=400$ and $\sigma=1$, the uniform kernel function for smoothing, and least-squares cross-validation bandwidth selection method$.$]{Fitting the isotropic and anisotropic local constant estimators to the continuous regression function with a jump data and comparing to the underlying data generating process (truth)$.~$This plot uses simulation parameters $n=400$ and $\sigma=1$, the uniform kernel function for smoothing, and least-squares cross-validation bandwidth selection method$.$}\label{fig:continuousWithJump}
\end{figure}

Table \ref{tbl:simulationResultsTableSigma1} shows the continuous data, except for a jump in the regression function. We see a similar result as when smoothing the piecewise constant regression function data, where the ALC estimator has a smaller MESE than the LC estimator. However, the smaller sample standard deviations of the MESE for the LC estimator over the ALC estimator in Table \ref{tbl:simulationResultsTableSigma1SD} suggest that these smooths may not be significantly different from each other. Figure \ref{fig:boxplotCJ} shows a box plot of MESEs for fitting the LC, ALC, and ALCT to data of a continuous regression function with a jump. The figure shows that ALC improves upon the LC estimator. Figure \ref{fig:continuousWithJump} shows a single fit of the kernel estimators to data of a continuous regression function with a jump. We see that for the ALC estimator, the improvement at the change-point is balanced by the loss of information in areas where $g'(x)$ is large, and the range kernel loses information local in the $X$ direction when there is no jump causing undersmoothing.

\subsection{Two-dimensional simulated fire spread data} \label{sec:twoDfiresim}
We have shown that for one-dimensional data, our anisotropic smoothing methods show substantial improvement of fit in the presence of change-points. We are interested in smoothing imagery data collected from fire spread experimentation. The data contains a two-dimensional change-point at the fuel and burning border. In this section, we demonstrate the effectiveness of using anisotropic diffusion filtering and smoothing on simulated data similar to measurements collected from fire smoldering experiments. 

Fire spread videos start with a sheet of fuel where all RGB channel values are high. When a fire begins, areas that are burning have a high red channel value, and low green and blue channel values. Burnt-out areas all have low RGB channel values. The simulated fire data reflects the red value of RGB channels, is calculated using the data generating process
\begin{eqnarray}
g_j(x_{1i},x_{2i})=\left\{ \begin{matrix} 80, & (x_{1i}-x_{1o})^2+(x_{2i}-x_{2o})^2 < r(t_j)^2, \\ 130, & \textup{otherwise,} \end{matrix}\right.
\end{eqnarray}
where the origin of the fire is at $(x_{1o},x_{2o})$, $t_j$ is $j^{\textup{th}}$ image frame and $r(t_j)$ is the radius of the fire front at frame $t_j$. The model is $Y_i(t_j)=Y_{ij}=g_j(x_{1i},x_{2i})+\epsilon_{ij}$, where the error is identically and independently normally distributed with mean 0 and variance $\sigma^2=20$. There are $M=15$ Monte Carlo replicates of one simulated fire over a 80$\times$80 pixel-grid. The choice for the variance is based the observed variability in experimental fire measurements. 

\begin{figure}[h!]
\centering
\begin{subfigure}{.5\textwidth}
  \centering
  \includegraphics[width=1.2\textwidth]{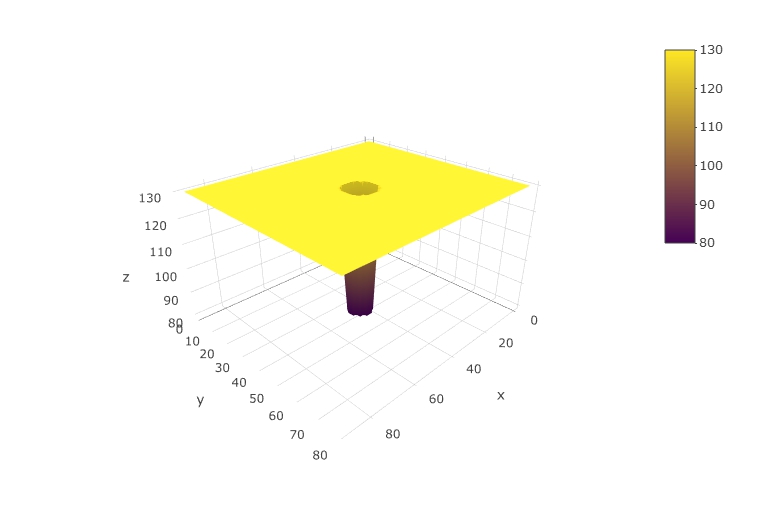}
  \caption{Data generating process at frame 10}
  \label{fig:simFireData1}
\end{subfigure}%
\begin{subfigure}{.5\textwidth}
  \centering
  \includegraphics[width=1.2\textwidth]{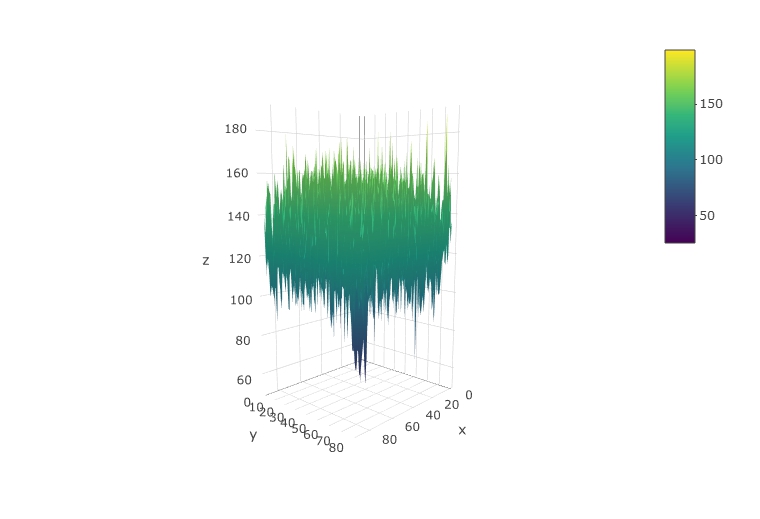}
  \caption{Frame 10 with noise}
  \label{fig:simFireData2}
\end{subfigure}
\begin{subfigure}{.5\textwidth}
  \centering
  \includegraphics[width=1.2\textwidth]{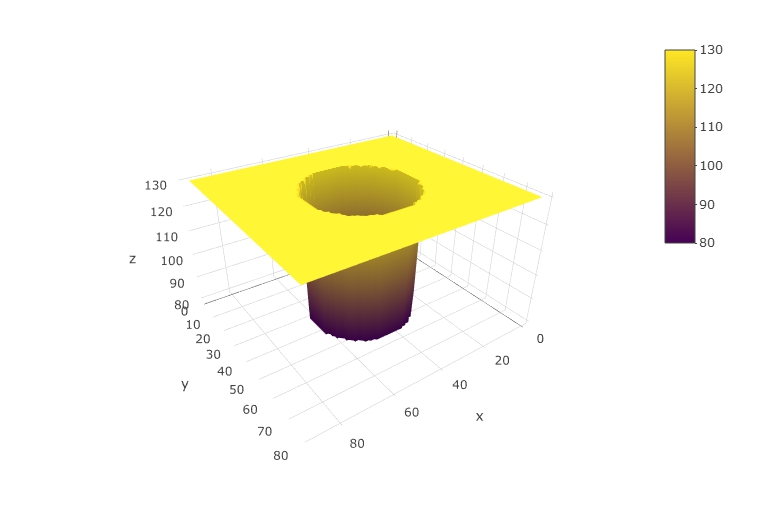}
  \caption{Data generating process at frame 35}
  \label{fig:simFireData3}
\end{subfigure}%
\begin{subfigure}{.5\textwidth}
  \centering
  \includegraphics[width=1.2\textwidth]{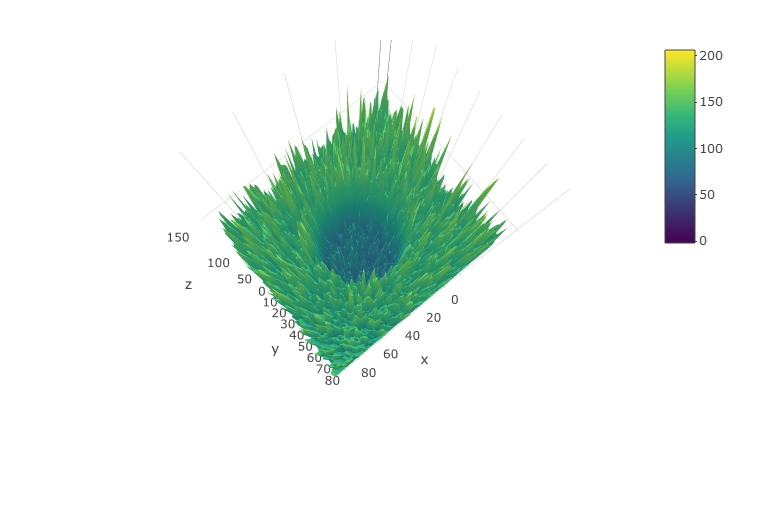}
  \caption{Frame 35 with noise}
  \label{fig:simFireData4}
\end{subfigure}
\begin{subfigure}{.5\textwidth}
  \centering
  \includegraphics[width=1.2\textwidth]{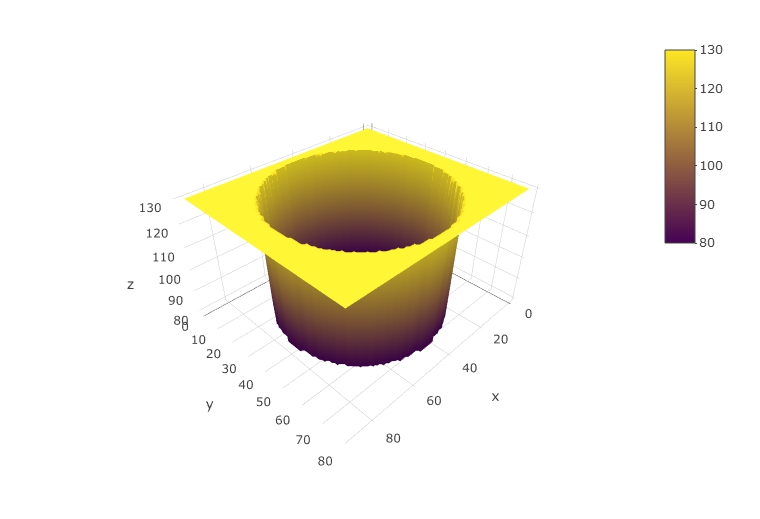}
  \caption{Data generating process at frame 60}
  \label{fig:simFireData5}
\end{subfigure}%
\begin{subfigure}{.5\textwidth}
  \centering
  \includegraphics[width=1.2\textwidth]{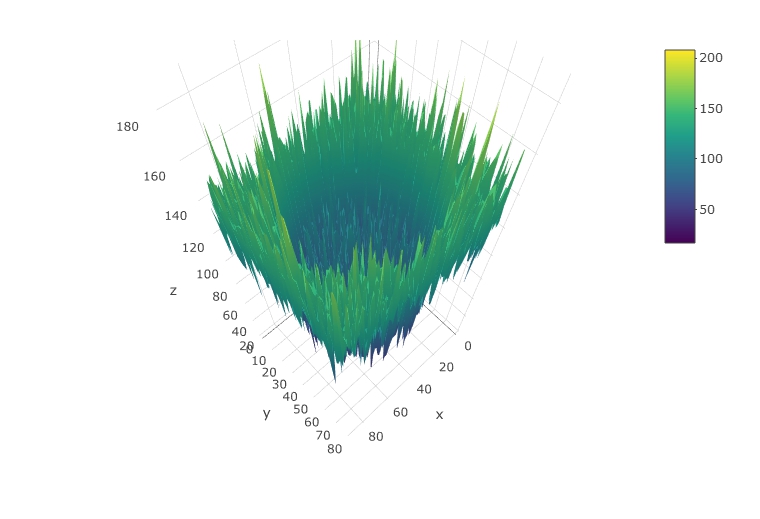}
  \caption{Frame 60 with noise}
  \label{fig:simFireData6}
\end{subfigure}
\caption{Examples of a simulated fire data generating process with and without noise. The simulation's video length is 70 equally spaced (in time) frames.}
\label{fig:simFireData}
\end{figure}

\begin{figure}[h!]
\centering
\begin{subfigure}{.5\textwidth}
  \centering
  \includegraphics[width=1.2\textwidth]{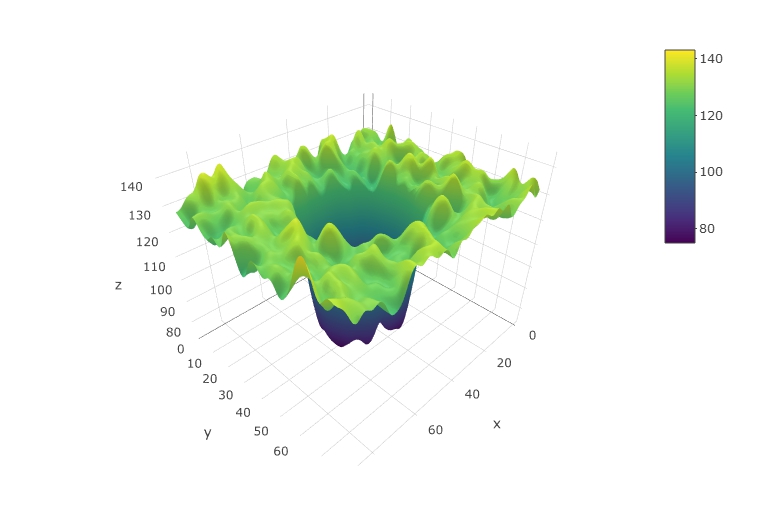}
  \caption{Isotropic local constant estimator}
  \label{fig:simFireDataFit35_2}
\end{subfigure}
\begin{subfigure}{.5\textwidth}
  \centering
  \includegraphics[width=1.2\textwidth]{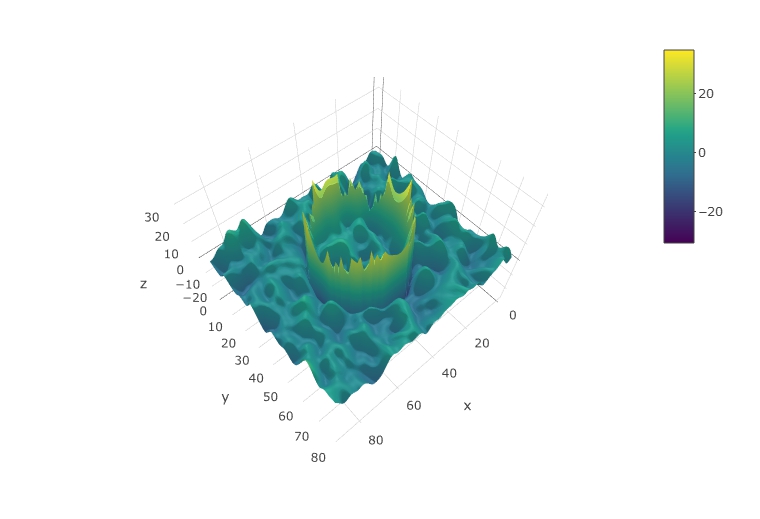}
  \caption{Isotropic minus data generating process}
  \label{fig:simFireDataFit35_3}
\end{subfigure}
\caption{The isotropic local constant estimator applied to image frame $t_{35}$ of the simulated fire spread data, and the estimated errors.}
\label{fig:simFireDataFit35_one}
\end{figure}

\begin{figure}[h!]
\centering
\begin{subfigure}{.5\textwidth}
  \centering
  \includegraphics[width=1.2\textwidth]{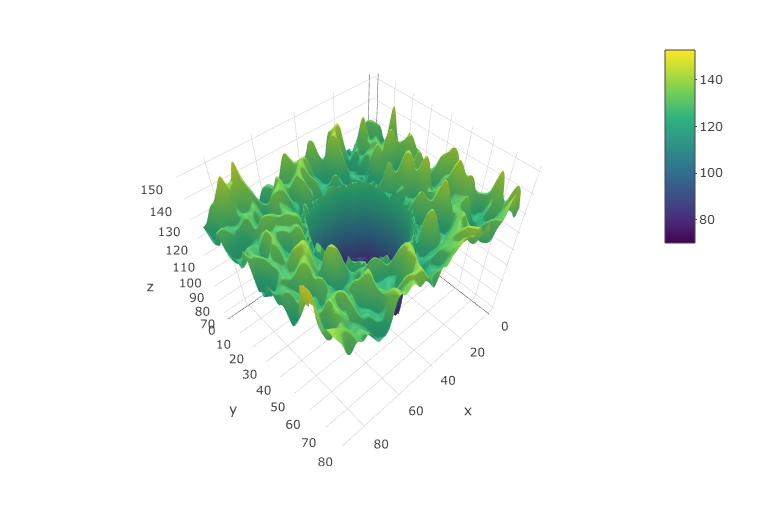}
  \caption{Anisotropic local constant estimator}
  \label{fig:simFireDataFit35_4}
\end{subfigure}%
\begin{subfigure}{.5\textwidth}
  \centering
  \includegraphics[width=1.2\textwidth]{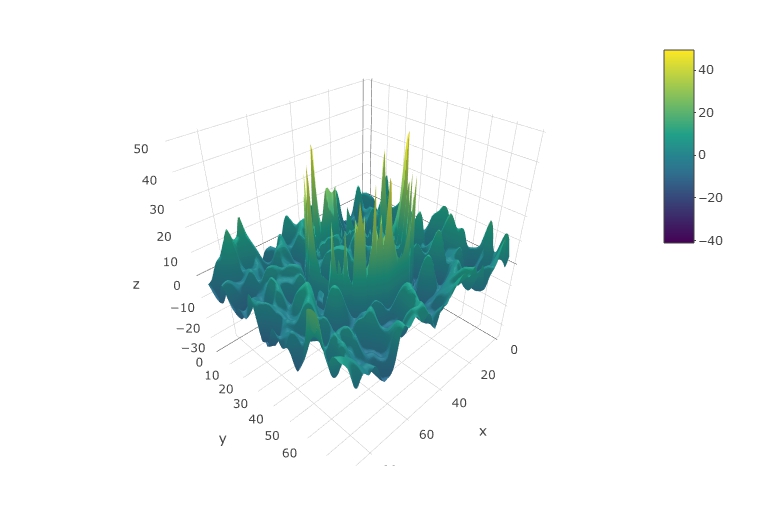}
  \caption{Anisotropic local constant estimator minus data generating process}
  \label{fig:simFireDataFit35_5}
\end{subfigure}
\begin{subfigure}{.5\textwidth}
  \centering
  \includegraphics[width=1.2\textwidth]{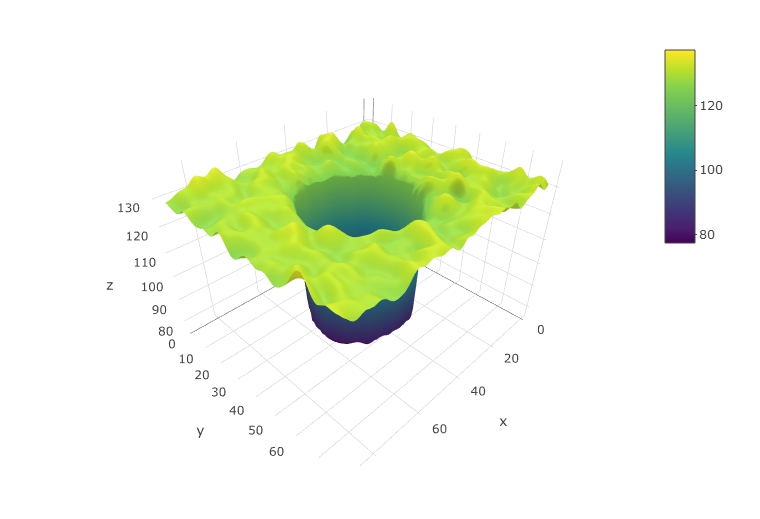}
  \caption{Oversmoothed anisotropic local constant estimator}
  \label{fig:simFireDataFit35_6}
\end{subfigure}%
\begin{subfigure}{.5\textwidth}
  \centering
  \includegraphics[width=1.2\textwidth]{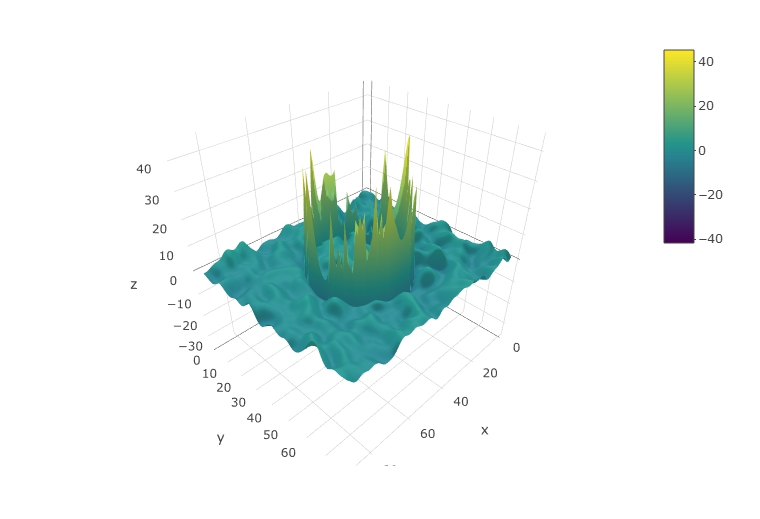}
  \caption{Oversmoothed anisotropic estimator minus data generating process}
  \label{fig:simFireDataFit35_7}
\end{subfigure}
\caption{The anisotropic local constant estimator applied to image frame $t_{35}$ of the simulated fire spread data, and the estimated errors. The first pair of plots use an ad hoc selected bandwidth. The second pair have the range kernel bandwidth increased by a factor of five.}
\label{fig:simFireDataFit35_two}
\end{figure}

The left column of Figure \ref{fig:simFireData} shows the data generating process, and the right column shows the data generating process with noise. The application of an isotropic local constant kernel estimator yields Figure \ref{fig:simFireDataFit35_2}, where there is relatively large error around the change-point, shown by a plot of the estimated errors in Figure \ref{fig:simFireDataFit35_3}. The two-dimensional anisotropic local constant kernel estimator is shown in Figure \ref{fig:simFireDataFit35_4}, where Figure \ref{fig:simFireDataFit35_5} shows the estimated error, with less error around the jump, but more error in the flat regions. Figure \ref{fig:simFireDataFit35_6} shows the same ALC smoother with the range bandwidth increased by a factor of five. Figure \ref{fig:simFireDataFit35_7} shows the estimated error of the ALC with an increased range kernel bandwidth. By increasing the bandwidth for the kernel in the $z$-direction, the regions between boundaries have a decrease in error while minimally affecting the error in estimating the change-point.

\subsection{Micro-fire experimental data}
\begin{figure}[!htbp]
\centering
\begin{subfigure}{.5\textwidth}
  \centering
\includegraphics[width=.9\textwidth]{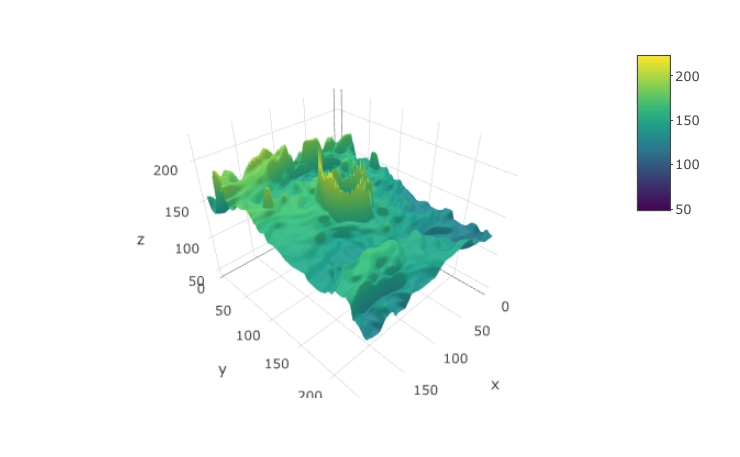}
  \caption{ALC smoothing}
  \label{fig:lcAndalcSmooth3}
\end{subfigure}%
\begin{subfigure}{.5\textwidth}
  \centering
\includegraphics[width=.9\textwidth]{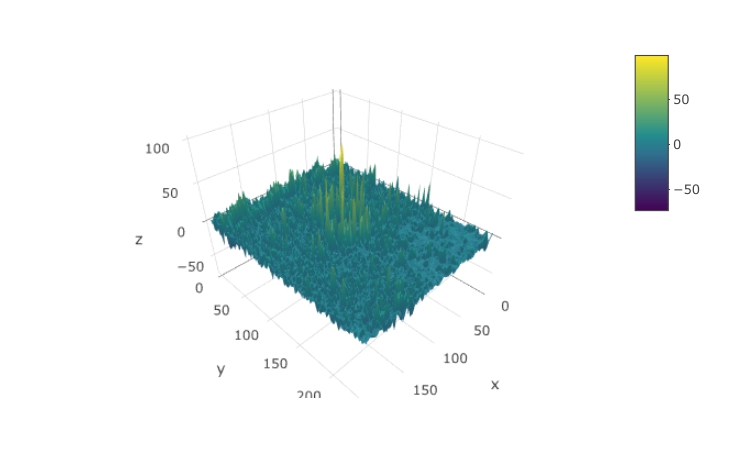}
  \caption{ALC residuals}
  \label{fig:lcAndalcSmooth4}
\end{subfigure}
\caption{The result for the red channel value after applying anisotropic local constant estimators on a smoldering image. The bandwidth for the anisotropic range kernel is selected using the ad hoc procedure. The $z$-axis is the RGB channel value, and $(x,y)$-directions are the pixel locations of the image.}
\label{fig:lcAndalcSmoothRed}
\end{figure}

\begin{figure}[!htbp]
\centering
\begin{subfigure}{.5\textwidth}
  \centering
\includegraphics[width=.9\textwidth]{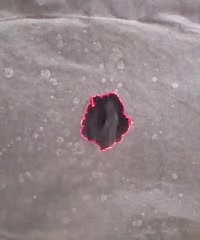}
  \caption{Original fire image}
  \label{fig:lcAndalcSmooth3}
\end{subfigure}%
\begin{subfigure}{.5\textwidth}
  \centering
\includegraphics[width=.9\textwidth]{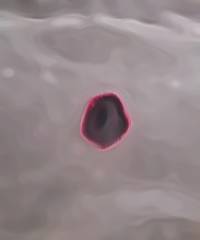}
  \caption{ALC smoothed image}
  \label{fig:lcAndalcSmooth4}
\end{subfigure}
\caption{The result of applying an anisotropic local constant estimator on a smoldering image. The bandwidth is selected using the AIC cross-validation method.}
\label{fig:lcAndalcSmooth}
\end{figure}

We consider the fire spread data from micro-fire experiments in \citep{thompson19}. Anisotropic smoothing is applied to fire spread measurements to preserve edges and smooth between the boundaries of fuel, burning, and burnt out regions. Figure \ref{fig:lcAndalcSmoothRed} shows the anisotropic local constant estimator applied to the red channel of the RGB fire spread images. This figure demonstrates the effectiveness of anisotropic smoothing, where the boundary is implicitly detected and not smoothed over. There are clear issues in the two regions that are separated by the fire boundary, where optimal bandwidth selection methods for isotropic local constant--such as least-squares cross-validation \citep{li07} or AIC cross-validation \citep{hurvich98}--do not perform optimally for anisotropic smoothing as they are designed for isotropic smoothing. Figure \ref{fig:lcAndalcSmooth} shows the resulting fire image from smoothing each of the three RGB channels independently. While the result is satisfactory in smoothing the areas between boundaries, the method needs improvement through an asymptotically optimal bandwidth selection procedure. 

\section{Discussion and conclusions}
There are a few areas where improvement is needed for the anisotropic smoothing framework, where the most important one is bandwidth selection. We show that theoretically we need relatively larger bandwidths for the anisotropic local constant estimator than the bandwidths used for the pilot isotropic or anisotropic local constant estimators. We show that using a slightly larger bandwidth than those chosen by our automatic bandwidth selectors for our two-dimensional simulated data estimator has the beneficial effect of decreasing the global mean of the MESE and thus improving the smooth. The development of a better bandwidth selector for single and iterative anisotropic smoothing could lead to improved estimates close to and remote from change-points, and in regions where $g'(x)/g(x)$ is large--an area where nonparametric estimators have been known to perform sub-optimally \citep{fan96}. Our current version of the estimator and bandwidth selection does not perform well in these regions, and undersmooths local information unlike the isotropic estimator.

The task of quantifying the efficiency of any number of iterative re-smooths requires investigation. Our current automatic bandwidth selections may be causing the iterative re-smoothing procedure to behave sub-optimally. We want to use iterative smoothing to improve $g(x)$ as best we can, but we don't want to perform any unnecessary iterative smooths that do not improve the fit, or cause under- or over-smoothing. We choose a conventional criterion to compare nonparametric smoothing methods: the MESE. Currently, the optimal number of iterative re-smooths is chosen by the minimum MESE of the iterations, which is generally after one anisotropic smoothing step. Therefore, we make the claim that minimum MSE or MESE may not be the most appropriate measure of an optimal smooth in the context of change-point regression function estimation. We believe there is a need to identify an optimal bandwidth selection method for anisotropic smoothing that balances accuracy around the change-point with accuracy farther away from the change-point, and this is future work.


Simulation sample sizes are deliberately small ($n=400$ for 1D, $n=80\times 80$ for 2D) to demonstrate the effectiveness of capturing change-points with less information. Larger simulation sizes were conducted (up to $n=128000$ for 1D), which showed similar results to those presented in this paper. We believe that the sample sizes shown here demonstrate the improvement in performance as sample size increases, particularly when using the underlying data generating process as the pilot estimator. We are interested in forest fire imagery applications that have low data resolution. Another issue is the number of Monte Carlo replicates; $M=125$ for one-dimensional data and $M=15$ for two-dimensional data. These step counts are quite small, however some simulations were run on larger number replicates where the results are similar.

Another interesting and subtle result arises from the rate of convergence of the anisotropic estimator. As shown in this chapter for an order $\nu=2$ kernel, the convergence rates for the isotropic and anisotropic local constant estimators are $O\left(n^{-\frac{2}{(q+4)}}\right)$ and $O\left(n^{-\frac{1}{(q+2)}}\right)$ respectively, where $q$ is the number of dimensions of the explanatory variables. The rate of convergence for a $q=1$ anisotropic local constant estimator is $O\left(n^{-\frac{1}{(3)}}\right)$, which is equivalent to the rate of convergence of a $q=2$ isotropic local constant estimator. This implies that adding the range kernel to the isotropic estimator is equivalent to adding another dimension to the problem. Similarly, the rate of convergence for a $q=2$ anisotropic local constant estimator is $O\left(n^{-\frac{1}{(5)}}\right)$, which is equivalent to the rate of convergence of a $q=4$ isotropic local constant estimator. Therefore, the underlying data generating process $g(X_i)$ that has dimension $q$ inputs into the range kernel is equivalent to doubling the number of dimensions in the rate of convergence of the isotropic estimator--that is $O\left(n^{-\frac{2}{(q+4)}}\right)$ becomes $O\left(n^{-\frac{2}{(2q+4)}}\right)=O\left(n^{-\frac{1}{(q+2)}}\right)$ when a range kernel is added that depends on $g(X_i)$.

In this paper, we have introduced a generalized form of an anisotropic local constant estimator that converges in probability to the underlying data generating process. We have shown the performance of the anisotropic local constant estimator with the local constant estimator as a pilot on one- and two-dimensional data. We believe that this estimator implicitly interprets jumps in the regression function as ``new boundaries" between regions. We applied the estimator to micro-fire spread imagery with some success. In the future, we look to improve the anisotropic local constant estimator as an image processor for experimental and real-world forest fire data. 

\section{Acknowledgements}

This research was supported, in part, by funding from the Natural Sciences and Engineering Research Council (NSERC) of Canada, Canadian Statistical Sciences Institute (CANSSI), and Institute for Catastrophic Loss Reduction (ICLR). The authors would like to thank X. Joey Wang (University of British Columbia, Okanagan), Douglas Woolford (University of Western Ontario), Charmaine Dean (University of Waterloo), and Mary Thompson (University of Waterloo) for their valuable input and insights. Lastly, we would like to thank the Shared Hierarchical Academic Research Computing Network (SHARCNET) for providing us with computational and educational resources used in simulations and analyses contained within this paper.

\bibliography{references}

\section{Appendix} \label{appendix}

\begin{table}[h!]
\centering
\caption{The mean of the MESEs for each nonparametric estimator fitted to the simulated piecewise constant regression function data.} 
\label{tbl:simulationResultsTableSigmaPoint1}
\scalebox{1}{
\begin{tabular}{|r|rrr|rrr|rrr|}
  \hline
 &   & $n=400$ &   &   & $n=800$ &   &   & $n=1600$ &   \\ 
  \hline
$\sigma$  & LC & ALC & ALCT & LC & ALC & ALCT & LC & ALC & ALCT \\ 
  0.1 & 0.00363 & 0.00179 & 0.00012 & 0.00336 & 0.00183 & 6e-05 & 0.00981 & 0.00019 & 3e-05 \\ 
  0.5 & 0.079 & 0.02119 & 0.003 & 0.05924 & 0.00868 & 0.00147 & 0.04283 & 0.00467 & 0.00084 \\ 
  1 & 0.1855 & 0.06835 & 0.01181 & 0.13443 & 0.04382 & 0.00579 & 0.10027 & 0.02887 & 0.00331 \\ 
  2 & 0.43075 & 0.24942 & 0.04743 & 0.29903 & 0.16317 & 0.02315 & 0.20681 & 0.10365 & 0.01318 \\ 
   \hline
\end{tabular}
}
\end{table}

\begin{table}[h!]
\centering
\caption{The sample standard deviation of the MESES for each nonparametric estimator fitted to the simulated piecewise constant regression function data.} 
\label{tbl:simulationResultsTableSigmaPoint1SD}
\scalebox{1}{
\begin{tabular}{|r|rrr|rrr|rrr|}
  \hline
 &   & $n=400$ &   &   & $n=800$ &   &   & $n=1600$ &   \\ 
  \hline
$\sigma$  & LC & ALC & ALCT & LC & ALC & ALCT & LC & ALC & ALCT \\ 
  0.1 & 0.00276 & 0.00105 & 0.00013 & 0.00027 & 0.00096 & 0.00006 & 0.00186 & 0.00059 & 0.00003 \\ 
  0.5 & 0.01993 & 0.01877 & 0.00316 & 0.00853 & 0.01187 & 0.00145 & 0.00298 & 0.00666 & 0.00085 \\ 
  1 & 0.03004 & 0.04527 & 0.0126 & 0.01816 & 0.03083 & 0.00577 & 0.01014 & 0.01661 & 0.00338 \\ 
  2 & 0.09884 & 0.12813 & 0.05201 & 0.04775 & 0.07455 & 0.023 & 0.0251 & 0.04209 & 0.01345 \\ 
   \hline
\end{tabular}
}
\end{table}

\begin{table}[h!]
\centering
\caption{The mean of the MESEs for each nonparametric estimator fitted to the simulated continuous regression function data.} 
\label{tbl:simulationResultsTableSigmaPoint5}
\scalebox{1}{
\begin{tabular}{|r|rrr|rrr|rrr|}
  \hline
 &   & $n=400$ &   &   & $n=800$ &   &   & $n=1600$ &   \\ 
  \hline
$\sigma$  & LC & ALC & ALCT & LC & ALC & ALCT & LC & ALC & ALCT \\ 
  0.1 & 0.00126 & 0.00189 & 0.00101 & 0.00074 & 0.00111 & 0.00057 & 0.00043 & 0.00065 & 0.00032 \\ 
  0.5 & 0.01473 & 0.01938 & 0.01239 & 0.00861 & 0.01176 & 0.00684 & 0.00513 & 0.00723 & 0.00414 \\ 
  1 & 0.0416 & 0.05359 & 0.03765 & 0.02364 & 0.03154 & 0.02078 & 0.01454 & 0.01959 & 0.01191 \\ 
  2 & 0.12214 & 0.15806 & 0.11457 & 0.0712 & 0.08885 & 0.06197 & 0.04204 & 0.05613 & 0.0363 \\ 
   \hline
\end{tabular}
}
\end{table}

\begin{table}[h!]
\centering
\caption{The sample standard deviation of the MESES for each nonparametric estimator fitted to the simulated continuous regression function data.} 
\label{tbl:simulationResultsTableSigmaPoint5SD}
\scalebox{1}{
\begin{tabular}{|r|rrr|rrr|rrr|}
  \hline
 &   & $n=400$ &   &   & $n=800$ &   &   & $n=1600$ &   \\ 
  \hline
$\sigma$  & LC & ALC & ALCT & LC & ALC & ALCT & LC & ALC & ALCT \\ 
  0.1 & 0.00031 & 0.00053 & 0.00025 & 0.00013 & 0.00026 & 0.00012 & 0.00007  & 0.00015 & 0.00007 \\ 
  0.5 & 0.00412 & 0.00525 & 0.00409 & 0.00257 & 0.00339 & 0.00225 & 0.00136 & 0.00211 & 0.00129 \\ 
  1 & 0.01616 & 0.01882 & 0.01604 & 0.00756 & 0.00941 & 0.00849 & 0.00454 & 0.00619 & 0.00476 \\ 
  2 & 0.06245 & 0.07175 & 0.06931 & 0.02959 & 0.03356 & 0.02964 & 0.01486 & 0.01943 & 0.01716 \\ 
   \hline
\end{tabular}
}
\end{table}

\begin{table}[h!]
\centering
\caption{The mean of the MESEs for each nonparametric estimator fitted to the simulated continuous data with a jump in the regression function.} 
\label{tbl:simulationResultsTableSigma1}
\scalebox{1}{
\begin{tabular}{|r|rrr|rrr|rrr|}
  \hline
 &   & $n=400$ &   &   & $n=800$ &   &   & $n=1600$ &   \\ 
  \hline
$\sigma$  & LC & ALC & ALCT & LC & ALC & ALCT & LC & ALC & ALCT \\ 
  0.1 & 0.01732 & 0.00319 & 0.00152 & 0.0106 & 0.00138 & 0.00087 & 0.00774 & 0.00104 & 0.00051 \\ 
  0.5 & 0.07338 & 0.04232 & 0.01516 & 0.05377 & 0.03069 & 0.00844 & 0.03689 & 0.02281 & 0.00503 \\ 
  1 & 0.15469 & 0.12098 & 0.04529 & 0.10683 & 0.07893 & 0.02545 & 0.07512 & 0.05144 & 0.01499 \\ 
  2 & 0.33026 & 0.28736 & 0.14387 & 0.2244 & 0.18895 & 0.07838 & 0.15711 & 0.12118 & 0.04594 \\ 
   \hline
\end{tabular}
}
\end{table}

\begin{table}[h!]
\centering
\caption{The sample standard deviation of the MESES for each nonparametric estimator fitted to the simulated continuous data with a jump in the regression function.} 
\label{tbl:simulationResultsTableSigma1SD}
\scalebox{1}{
\begin{tabular}{|r|rrr|rrr|rrr|}
  \hline
 &   & $n=400$ &   &   & $n=800$ &   &   & $n=1600$ &   \\ 
  \hline
$\sigma$  & LC & ALC & ALCT & LC & ALC & ALCT & LC & ALC & ALCT \\ 
  0.1 & 0.00109 & 0.00043 & 0.00030 & 0.00123 & 0.00033 & 0.00015 & 0.00071 & 0.00118 & 0.00008 \\ 
  0.5 & 0.00735 & 0.01848 & 0.005 & 0.00494 & 0.00995 & 0.00222 & 0.00281 & 0.00597 & 0.00155 \\ 
  1 & 0.03002 & 0.03732 & 0.01904 & 0.01328 & 0.01753 & 0.00843 & 0.00816 & 0.00956 & 0.00515 \\ 
  2 & 0.08068 & 0.08431 & 0.07131 & 0.03541 & 0.04431 & 0.02913 & 0.02699 & 0.02902 & 0.01662 \\ 
   \hline
\end{tabular}
}
\end{table}

\end{document}